\pdfoutput=1

\documentclass[%
 reprint,
 floatfix,
 widetext,
 amsmath,amssymb,
 aps,
]{revtex4-2}

\usepackage[strict]{changepage}
\usepackage[title]{appendix}
\usepackage{xcolor}
\usepackage{graphicx}
\usepackage{dcolumn}
\usepackage{bm}

\usepackage[utf8]{inputenc}
\usepackage{physics}
\usepackage{amsfonts}
\usepackage{amsmath}
\usepackage{comment}
\usepackage{float}

\begin{document}

\preprint{APS/123-QED}

\title{Mean-field approximation of the Hubbard model expressed in a many-body basis.}

\author{Antoine Honet}
\affiliation{%
Department of Physics and Namur Institute of Structured Materials, University of Namur, Rue de Bruxelles 51, 5000 Namur, Belgium
}%

\author{Luc Henrard}
\affiliation{%
Department of Physics and Namur Institute of Structured Materials, University of Namur, Rue de Bruxelles 51, 5000 Namur, Belgium
}%

\author{Vincent Meunier}%
\affiliation{%
Department of Engineering Science and Mechanics, The Pennsylvania State University, University Park, PA, USA
}%

\date{\today}

\begin{abstract}
 The effective independent-particle (mean-field) approximation of the Hubbard Hamiltonian is described in a many-body basis to develop a formal comparison with the exact diagonalization of the full Hubbard model, using small atomic chain as test systems. This allows for the development of an intuitive understanding of the shortcomings of the mean-field approximation and of how critical correlation effects are missed in this popular approach. The description in the many-body basis highlights a potential ambiguity related to the definition of the density of states. Specifically,  satellite peaks are shown to emerge in the mean-field approximation, in departure from the common belief that they characterize correlation effects. The scheme emphasizes the importance of correlation and how different many-body corrections can improve the mean-field description. The pedagogical treatment is expected to make it possible for researchers to acquire an improved understanding of many-body effects as found in various areas related to electronic properties of molecules and solids, which is highly relevant to current efforts in quantum information and quantum computing. 
 
\begin{description}
\item[Keywords]
Hubbard model, mean-field approximation, exact diagonalization, many-electron basis
\end{description}
\end{abstract}

\maketitle


\author{Antoine Honet, Luc Henrard and Vincent Meunier}

\section{Introduction}

The Hubbard model~\cite{hubbard_electron_1963} is a popular and simple model to describe electron correlation in solids, molecules, and nanoparticles. The exact description of correlation is a tremendous task in the field of electronics, and it is of paramount importance for the accurate description of magnetism, optical properties, electron transport, and plasmonics~\cite{yazyev_emergence_2010, bullard_improved_2015, lauchner_molecular_2015, thongrattanasiri_quantum_2012}.

In spite of its apparent simplicity, finding exact solutions of the Hubbard model is a formidable effort in general. Starting from the Hubbard Hamiltonian, the simplest conceptual way to solve it is the exact diagonalization (ED) method~\cite{jafari_introduction_2008, kingsley_exact_2013, sharma_organization_2015}, that can however only be performed analytically or numerically for very small systems~\cite{romaniello_self-energy_2009, romaniello_beyond_2012, strunck_combining_nodate, tomczak_proprietes_2007, yepez_lecture_nodate, eder_introduction_2017, jafari_introduction_2008, kingsley_exact_2013, sharma_organization_2015}. This method hinges on the diagonalization of the Hubbard Hamiltonian expressed in a many-electron basis and yields the eigen-energies and eigen-vectors of the Hamiltonian, expressed as linear combinations of the (many-body) basis states. The major issue with this method is that the number of basis states grows exponentially with the number of electrons. For example, at half-filling, the many-body basis for the single-orbital Hubbard model of a two-site system has dimension $6$, dimension $20$ for three-site system, and already dimension $924$ for a six-site system. The dimension of the Hilbert space for a ten-site system reaches $184,756$, leading to a Hamiltonian matrix with more that $3.4 \cross 10^{10}$ elements~\cite{kingsley_exact_2013}. This illustrates how the numerical resolution of the method is rapidly limited, even for modest size systems. For this reason, researchers have realized the need for approximation methods to render the computational treatment of the Hubbard model at a tractable computational cost.

The mean-field approximation (MF) is one of the simplest approximations for the Hubbard model. It consists in replacing the two-body interaction term of the Hubbard model as an interaction between one electron and a mean-field due to the other electrons. As a result, a given electron no longer interacts directly with other electrons but rather with a field. This makes it possible to write the Hamiltonian in a single-electron basis and to consider electrons as particles that are effectively free. The main advantage of MF is that the corresponding dimension of the Hilbert space is reduced to $2N_{\rm el}$, where $N_{\rm el}$ is the number of electrons and the factor $2$ accounts for the spin degree of freedom. The MF approximation is often used to describe electronic systems~\cite{yazyev_emergence_2010, bullard_improved_2015, feldner_magnetism_2010} and we will refer to the formulation of the MF approximation in the single-electron basis as MF-U for \textit{mean-field usual}.

One major drawback of the MF-U approximation is that it misses all the correlation between electrons. Several techniques have been developed to move beyond MF-U to include (at least a part of the) correlation, such as the Green's function many-body approximation (GFMBA), which consists of a sum of Feynman diagrams~\cite{stefanucci_nonequilibrium_2013, schlunzen_nonequilibrium_2016}. This family of approximations includes the second-order Born approximation, the GW approximation, and the T-matrix approximation. There exists several other approaches such as the dynamical mean-field theory and the quantum Monte Carlo approaches.~\cite{feldner_magnetism_2010, raczkowski_hubbard_2020}. Furthermore, other methods are based on the MF-U computation augmented by a symmetry restoration procedure~\cite{yannouleas_symmetry_2007, sheikh_symmetry_2021}. What's more, machine learning based self-energy construction have also been investigated more recently~\cite{tirimbo_kernel_nodate, sandberg_machine-learning_nodate, song_analytic_2020}.

Up to now, the discrepancies between MF-U and ED wave functions have not been fully described or understood in a general framework despite their importance for the development of intuitive and accurate corrections to the approximation. We believe that this lack of a deeper understanding is partly due to the fact that ED methods must be expressed in the many-electron basis whereas the MF-U is, by design, usually implemented in the single-electron basis. It is therefore challenging to compare the two methods since this change of basis is not a simple unitary transformation but a complete change of paradigm. 

The objective of this paper is to propose an in-depth comparison between MF and ED by formulating the MF approximation in the many-body basis framework. This approximation will be referred to as the MF-MB in the rest of the paper. The goal of this formulation is not to reduce the computational cost of the method but to build a MF approximation in the same basis as the ED, and thus to gain insight into the missing part of the MF approximation (\textit{i.e.}, correlation). We note, however, that in contrast to common density functional theory computations, the Hubbard MF approximation does not suffer from an \textit{exchange problem} since the ground state is approximated by a Slater determinant that satisfies the wave function symmetry required for the eigen-states of the exchange operator, which commutes with the Hamiltonian of the electronic systems.

Our analysis highlights the ambiguous definition of the density of states within the MF approximation depending on the method. The MF-U method is often combined implicitly with Koopman's theorem~\cite{koopmans_uber_1934, szabo_modern_1996}, which assumes that all the states of the $N$-electron system are frozen when adding/removing an electron. This assumption is not strictly correct since even in the MF approximation, adding or removing an electron changes the mean-field and consequently the predicted states (both occupied and unoccupied) as we will illustrate by comparing the DOS obtained \textit{via} MF-U and MF-MB techniques.  This comparison highlights the fact that satellite states also appear in the MF-MB approximation, although they are often thought as being the result of the inclusion of correlation~\cite{reining_gw_2018}.

The rest of the paper is organized as follow: we first introduce the Hubbard model, the MF-U approximation, and the ED technique. We then explain the MF-MB technique, based both on MF-U equations and the numerical methods employed for the ED. Finally, we discuss results of MF-MB compared with MF-U and ED and the notion of density of states and the appearance of satellite peaks in the MF-MB.

\section{Review of standard methods}

\subsection{Hubbard model}

This studies focused on the single-orbital Fermi-Hubbard model:
\begin{equation}
\hat{H}_{\rm Hubbard} = - t \sum_{<ij>, \sigma}  \hat{c}^\dagger_{i\sigma} \hat{c}_{j\sigma}  + U \sum_i \hat{n}_{i \uparrow} \hat{n}_{i \downarrow}
\label{Hubbard_ham}
\end{equation}
where $t$ is the hopping parameter, $U$ is the interaction (or Hubbard) parameter, $\hat{c}^\dagger_{i\sigma}$ (resp. $\hat{c}_{i\sigma}$) is a creation (resp. destruction) operator of an electron on atomic site $i$ with spin $\sigma$, and $\hat{n}_{i \sigma} = \hat{c}^\dagger_{i\sigma} \hat{c}_{i\sigma}$ is the density operator (on atomic site $i$ and with spin $\sigma$). The atomic site indices run from $0$ to $N-1$ where $N$ is the total number of sites. The $\langle~.~\rangle$ symbol under the summation operator indicates that the sum runs over all pairs of nearest-neighbour atomic sites.

The first term of the Hubbard Hamiltonian in eq.~\ref{Hubbard_ham} is the tight-binding Hamiltonian and is easily written using the one-electron basis. In contrast, the second term (known as the Hubbard or interaction term) is the product of two density operators (\textit{i.e.}, a combination of four creation and/or destruction operators). This two-body operator cannot be written in the one-electron basis.

\subsection{Mean-field approximation and single-electron basis}

\label{sec:MF-U}

In the MF approximation, the Hubbard Hamiltonian of eq.~(\ref{Hubbard_ham}) is approximated so that it only includes one-body operators and the Hamiltonian can be written in the one-electron basis. In practice, the density operators are decomposed as sums of the mean value of the operator ($n_{i \sigma}$) and the deviation ($\hat{n}_{i \sigma} - n_{i \sigma}$) from this mean value: $\hat{n}_{i \sigma} = n_{i \sigma} + (\hat{n}_{i \sigma} - n_{i \sigma})$. Products of density operators in eq.~(\ref{Hubbard_ham}) are expanded and the approximation consists in dropping products of deviations from the mean, generally assumed (without proof) to be small. Products of mean values only induce a constant shift in the Hamiltonian (and thus in the total energy), leading to the Hamiltonian:
\begin{equation}
\begin{split}
   \hat{H}_{Hubb,MF} = & - t \sum_{<ij>, \sigma}  \hat{c}^\dagger_{i\sigma} \hat{c}_{j\sigma}   + U \sum_i ( n_{i \uparrow} \hat{n}_{i \downarrow} + n_{i \downarrow} \hat{n}_{i \uparrow}) \\
   &-  U \sum_i n_{i \uparrow} n_{i \downarrow}.
\end{split}
\label{eq:ham_hubb_mf}
\end{equation}

\begin{figure}
\centering
    \includegraphics[width=9cm]{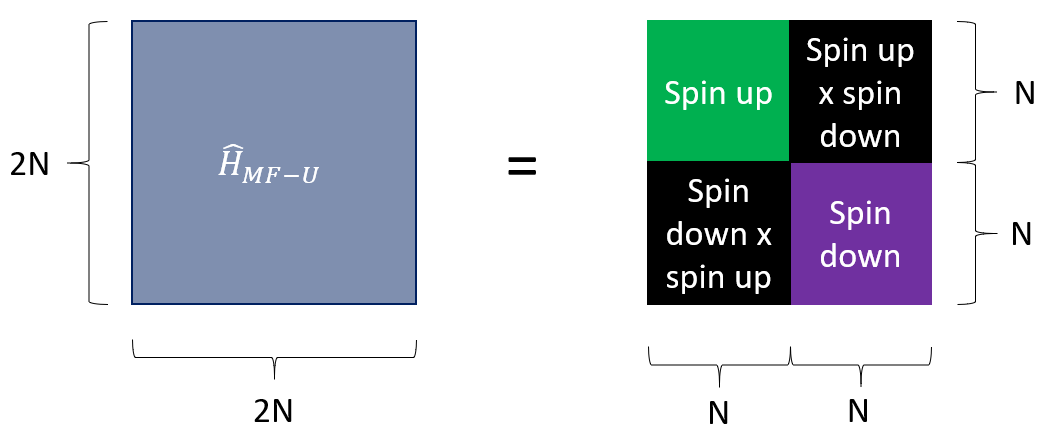}
\caption{Illustration of the structure of the Hubbard Hamiltonian in the MF-U approximation. The Hamiltonian matrix of size $2N \times 2N$ is written in the single-electron basis and can be divided into $4$ blocks of size $N\times N$: one containing pure spin-up related terms, one containing pure spin-down terms, and two mixing blocks between up and down spins.}
\label{fig:illustration_2Nx2N}
\end{figure}

In the MF-U method, the Hamiltonian of eq.~(\ref{eq:ham_hubb_mf}) is written in the form of a matrix expressed in the one-electron basis. The basis states of the one-electron basis are obtained by the application of the different creation operators on the vacuum state $\ket{\emptyset}$ (\textit{i.e.}, the state containing no electron), for each site and spin value. It follows that 
\begin{equation}
    \ket{i\sigma} = \hat{c}^\dagger_{i\sigma} \ket{\emptyset}
    \label{eq:single_electron_basis}
\end{equation}
are the $2N$ basis states of the single-electron basis. In that basis, the matrix has dimension $2N \times 2N$ ($N$ being the number of sites). The Hamiltonian matrix is composed of $4$ blocks of dimension $N\times N$: the two blocks on the diagonal are pure spin-up and spin-down blocks and the two off-diagonal blocks mix spin up and spin down that, according to the MF Hamiltonian, remain equal to zero (see figure~\ref{fig:illustration_2Nx2N} for an illustration of the block matrix). The tight-binding term implies that the Hamiltonian matrix has terms of amplitude $-t$ for elements corresponding to neighbouring sites. The second term of the Hamiltonian is concerned with diagonal elements in the single-electron basis. The diagonal elements in the spin-up (-down) block involve mean densities of down (up) spins. One usually does not implement the last term (which only includes mean values, not operators) in eq.~(\ref{eq:ham_hubb_mf}). We need, however, to reintroduce that term in the computation of the total energy, as it corresponds to a rigid shift of all energy levels.

For example, the mean-field Hubbard Hamiltonian (eq.~(\ref{eq:ham_hubb_mf})) is expressed in the single-electron basis for the two-site system as:
\begin{equation}
    \hat{H}_{Hubb,\textrm{MF-U}} = \mqty( U n_{0\downarrow} & -t & 0  & 0 \\ -t  & U n_{1\downarrow} & 0 & 0 \\ 0 & 0 & U n_{0\uparrow} & -t \\ 0 & 0 & -t & U n_{1\uparrow}  ), 
\end{equation}
if the basis states are ordered in spin blocks in the same way as atomic sites $0$ and $1$: $\{\ket{0\uparrow},\ket{1\uparrow},\ket{0\downarrow},\ket{1\downarrow}\}$.

The Hamiltonian of eq.~(\ref{eq:ham_hubb_mf}) involves mean values of density operators and it is therefore necessary to solve it self-consistently. Initial conditions for the mean densities are guessed and a Hamiltonian based on these mean values is built. In the next step, the first Hamiltonian is diagonalized, resulting in eigen-energies and eigen-vectors. New mean values are computed by populating the eigen-vectors of lowest energies and a new Hamiltonian is generated, diagonalized, leading to new mean values. This loop is repeated until the difference between the input mean values and the output values is smaller than a given threshold (convergence). 

The density of states (DOS) in the MF-U is constructed as a sum of Dirac delta-peaks centered at the converged eigen-energies of the Hamiltonian. A single-particle states is identified at each eigen-energy.

In addition, using the single electronic eigen-energies $E_k$ and the coefficients  $a_{i\sigma}^k$ of the eigen-states of the MF-U Hamiltonian ($k$ ranging from $0$ to $2N$), we define creation operators of each eigen-state as: 
\begin{equation}
    \hat{d}^\dagger_{k} = \sum_{i, \sigma} (a_{i\sigma}^k)^* \hat{c}^\dagger_{i\sigma}.
\end{equation}

For a system with $N_{\rm el}$ electrons, the wave-function of the MF-U ground state is expressed as a single Slater determinant, created by applying successively the $N_{\rm el}$ creation operators associated with the $N_{\rm el}$ lowest eigen-energies to the vacuum state:
\begin{equation}
    \ket{GS,\textrm{MF-U}} = \prod_{k \le N_{\rm el}} \hat{d}^\dagger_{k} \ket{\emptyset}
\end{equation}
and the mean values of density operators to be inserted in the self-consistent Hamiltonian are given by:
\begin{equation}
    n_{i\sigma} = \sum_{k\le N_{\rm el}} \abs{a_{i\sigma}^k}^2.
\end{equation}

\subsection{Exact diagonalization}
\label{sec:ED}

Instead of solving for the approximated MF Hamiltonian described in eq.~(\ref{eq:ham_hubb_mf}), the ED considers the exact Hubbard Hamiltonian of eq.~(\ref{Hubbard_ham}). The second term of the Hamiltonian, \textit{i.e.}, the product of two density operators, cannot be expressed in the single-electron basis of eq.~(\ref{eq:single_electron_basis}) and a many-electron basis needs to be used. As a common practice in the ED literature, the concept of \textit{sector} is introduced for a fixed number of electrons. When restricted to the $N_{\rm el}$ sector, the many-electron basis states are given by all possible combinations of $N_{\rm el}$ creation operators applied to the vacuum state. For example
\begin{equation}
    \prod_{k\le N_{\rm el}} \hat{c}^\dagger_{k\uparrow} \ket{\emptyset}
    \label{eq:many_el_basis}
\end{equation}
represents one of the basis states, involving only spin-up electrons.

As fermionic creation operators with at least one different index (site or spin) anti-commute, the order in the product only affects the sign of the state (\textit{i.e.}, the \textit{fermionic sign}) and we find the same basis state, \textit{modulo} an overall phase. Here, we conventionally choose to define the basis states with a positive sign by ordering all spin-up (resp. -down) creation operators to the left (resp. right); operators with the same spin part are ordered from left to right in an increasing order. This results in the combination formula in combinatorics for the number of basis states ($N_{b}$):
\begin{equation}
    N_b = C^{2N}_{N_{\rm el}} = \frac{(2N) !}{(2N-N_{\rm el})! \hspace{0.1cm} N_{\rm el}!} .
\end{equation}

We adopt the approach described in Refs.~\onlinecite{lin_exact_1993, jafari_introduction_2008,kingsley_exact_2013, sharma_organization_2015} to label each state with one integer $I$, bijectively linked with two other integers $I_{\uparrow}$ and $I_{\downarrow}$ by the relations:
\begin{equation}
I=2^N I_{\uparrow} + I_{\downarrow}
\end{equation}
and
\begin{equation}
\begin{split}
&I_{\uparrow} = I//2^N \\
&I_{\downarrow} = I \mod{2^N}, 
\end{split}
\label{eq:relation_I_Iup_Idown}
\end{equation}
where $//$ represents the integer division.

Writing $I_{\uparrow}$ (resp. $I_{\downarrow}$) in binary notation yields the space configuration of the state in the spin-up (resp. -down) sector. Organizing the Hilbert space in this way allows one to only have to deal with integers and to easily find the effect of the creation, destruction, and density operators on each state using simple standard binary operations (\textit{e.g.}, bin flip, bin counting,\ldots). One can now  express the action of the full Hubbard Hamiltonian of eq.~(\ref{Hubbard_ham}) on the basis states and, in turn, calculate the Hamiltonian matrix elements in the many-electron basis. 

Formally, the Hamiltonian matrix contains elements with value $-t$ when they correspond to connected basis states having all the same electron creation operators (sites and spin) but one. The different electrons have to be of the same spin on a neighboring atomic site. The Hamiltonian matrix also contains elements of amplitude $U$ on the diagonal for basis states containing two electrons on the same site and of opposite spin. The $U$ values are added if there are several doubly-occupied sites in the basis state. 

We illustrate the construction of the Hamiltonian for a two-site system at half-filling. The $6$ basis states are $\ket{\Phi_1} = \hat{c}^\dagger_{0, \uparrow} \hat{c}^\dagger_{1, \uparrow} \ket{\emptyset}, \ket{\Phi_2} = \hat{c}^\dagger_{0, \uparrow} \hat{c}^\dagger_{1, \downarrow} \ket{\emptyset}, \ket{\Phi_3} = \hat{c}^\dagger_{1, \uparrow} \hat{c}^\dagger_{0, \downarrow} \ket{\emptyset}, \ket{\Phi_4}  = \hat{c}^\dagger_{0, \uparrow} \hat{c}^\dagger_{0, \downarrow} \ket{\emptyset}, \ket{\Phi_5} =  \hat{c}^\dagger_{1, \uparrow} \hat{c}^\dagger_{1, \downarrow} \ket{\emptyset} $ and $\ket{\Phi_6} =  \hat{c}^\dagger_{0, \downarrow} \hat{c}^\dagger_{1, \downarrow} \ket{\emptyset}$ and the Hamiltonian matrix in that basis is given by:
\begin{equation}
H_{Hubb, ED} = \mqty(0 &  0 & 0 & 0 &  0 & 0 \\ 0 &  0 & 0 & -t &  -t & 0 \\  0 &  0 & 0 & -t &  -t & 0 \\  0 &  -t & -t & U &  0 & 0 \\  0 &  -t & -t & 0 &  U & 0 \\ 0 &  0 & 0 & 0 &  0 & 0 ),
\label{eq:H_hub_many_el_basis}
\end{equation}
where the element $-t$ connects basis states $\ket{\Phi_2}$ and $\ket{\Phi_3}$ to basis states $\ket{\Phi_4}$ and $\ket{\Phi_5}$ since they differ only by one electron having the same spin and hopping from site $0$ to $1$. $U$ is on the diagonal for basis states $\ket{\Phi_4}$ and $\ket{\Phi_5}$ both having two electrons, located on site $0$ and site $1$, respectively.

The lowest eigen-energy of the Hamiltonian corresponds to the ground-state and the higher ones to excited states. The corresponding eigen-vectors are by construction many-electron states. In contrast to the single-electron basis formulation of the MF-U, it is not constructed by populating several eigen-vectors. Likewise, the associated eigen-energy is the total energy of the system, without having to sum individual electron energies.

To gain access to dynamical (\textit{i.e.}, frequency dependent) properties, we now introduce the Green's function with general definition~\cite{stefanucci_nonequilibrium_2013,pollehn_assessment_1998, schindlmayr_spectra_nodate}:
\begin{equation}
\begin{split}
G_{i\sigma, j\sigma'} (\omega) = & \bra{\Psi_0^{N_{\rm el}}} \hat{c}_{i\sigma} \frac{1} {\omega + (E_0^{N_{\rm el}}-\hat{H}_{N_{\rm el}+1} +i\eta)} \hat{c}^\dagger_{j\sigma'} \ket{\Psi_0^{N_{\rm el}}} \\
& + \bra{\Psi_0^{N_{\rm el}}} \hat{c}^\dagger_{i\sigma} \frac{1} {\omega - (E_0^{N_{\rm el}}-\hat{H}_{N_{\rm el}-1}+i\eta)} \hat{c}_{j\sigma'} \ket{\Psi_0^{N_{\rm el}}},
\end{split}
\label{eq:many_body_greens_function}
\end{equation}
where $\ket{\Psi_0^{N_{\rm el}}}$ and $E_0^{N_{\rm el}}$ are the ground state and the ground-state energy of the system with $N_{\rm el}$ electrons and $\hat{H}_{N_{\rm el}\pm 1}$ are the Hamiltonian operators of the system containing ${N_{\rm el}\pm 1}$ electrons and $\eta$ is a small real positive parameter.

The first term in eq.~(\ref{eq:many_body_greens_function}) is the electron addition part of the Green's function: $\hat{c}^\dagger_{j\sigma'} \ket{\Psi_0^{N_{\rm el}}}$ and $\bra{\Psi_0^{N_{\rm el}}} \hat{c}_{i\sigma}$ represent both states with $N_{\rm el}+1$ electrons and the operator involves the $N_{\rm el}+1$ electron Hamiltonian. This term thus explores the possible states when an electron is added to the $N_{\rm el}$ ground state. The second term of eq.~(\ref{eq:many_body_greens_function}) describes the situation where one electron is removed. This is the electron removal part of the Green's function. The Green's function has poles at the frequencies corresponding to difference of energies between the $N_{\rm el}$ ground state and states in the $N_{\rm el} \pm 1$ sectors.

To evaluate the Green's function~(eq. (\ref{eq:many_body_greens_function})), three exact diagonalizations are completed: one in the $N_{\rm el}$ sector to find the ground-state $\ket{\Psi_0^{N_{\rm el}}}$ and its energy $E_0^{N_{\rm el}}$, and two for the two sectors $N_{\rm el}\pm1$ so that matrix elements of the type $\bra{\Phi_k^{N_{\rm el}\pm1}} \hat{H}_{N_{\rm el}\pm1} \ket{\Phi_{k'}^{N_{\rm el}\pm1}}$ can be computed with $\ket{\Phi_{k'}^{N_{\rm el}\pm1}}$ the basis vectors of the $N_{\rm el}\pm1$ sector of the Hilbert space. The Green's function is then computed by writing the states $\hat{c}^\dagger_{j\sigma'} \ket{\Psi_0^{N_{\rm el}}}$ in the basis $\ket{\Phi_{k}^{N_{\rm el}+1}}$ and the states $\hat{c}_{j\sigma'} \ket{\Psi_0^{N_{\rm el}}}$ in the basis $\ket{\Phi_{k}^{N_{\rm el}-1}}$. 

The DOS ($D (\omega)$) is defined from the Green's function as:
\begin{equation}
D (\omega) = -\frac{1}{\pi} {\rm Tr} \bigg( \Im(G^R(\omega)) \bigg),
\label{eq:DOS_MB_basis}
\end{equation}
where ${\rm Tr}$ is the trace.

We can now identify addition and removal parts of the density of states: the addition part is computed based on the addition part of the Green's function and, in an analogous way, the removal part of the DOS is constructed from the removal part of the Green's function. The DOS features peaks at energies corresponding to the poles of the Green's function, \textit{i.e.}, the differences between the $N_{\rm el}$ ground-state energy and all states (ground state and excited states) of $N_{\rm el} \pm 1$. We can imagine this process as probing all the possible \textit{adding} or \textit{removing} of energy when adding or removing one electron, taking into account the correlation exactly in both the starting state (ground state of $N_{\rm el}$ electron) and the final states ($N_{\rm el} \pm 1$ electron states). We point out that to account for the correlation, the $N_{\rm el} \pm 1$ states in the $N_{\rm el} \pm 1$ sectors cannot be obtained simply by adding one electron independently from the others.

\section{Mean-field approximation in the many-body basis}

\subsection{Formulation in the $N_{\rm el}$ sector}
As mentioned before, the MF approximation effectively decouples the density operators interaction into the interaction of one density operator with a mean-field. This results in the possible formulation within the single-electron basis of the MF approximation (see section~\ref{sec:MF-U}), reducing the basis dimension from $C^{2N}_{N_{\rm el}}$ to $2N$. We now examine how the MF approximation can be expressed in the many-body basis (MF-MB).

We consider the MF-approximated Hamiltonian of eq.~(\ref{eq:ham_hubb_mf}) and the many-electron basis described in eq.~(\ref{eq:many_el_basis}). As the MF approximation does not affect the tight-binding term of the Hamiltonian, the $-t$ elements of the MF-MB Hamiltonian matrix are the same as for the ED. In contrast, the interaction term involves mean values of density operators and the diagonal in the MF-MB Hamiltonian includes $U n_{i\sigma}$ factors for each basis states involving a creation operator $c^\dagger_{i\sigma'}$, $\sigma$ being one spin (up or down) and $\sigma'$ being the opposite spin. In general, there will be a sum of several $U n_{i\sigma}$ terms because basis states contain several electrons. For example, if there is a doubly-occupied site $i$ in a basis state, terms of the form $U (n_{i\sigma'} +n_{i\sigma}) $ are present on the diagonal of the Hamiltonian matrix.

We illustrate the construction of the Hamiltonian for the two-site system at half-filling with the basis states explicitly written in section~\ref{sec:ED}. The Hamiltonian matrix in this case is given by:

\begin{widetext}
\begin{equation}
H_{Hubb, \textrm{MF-MB}} = \mqty(U ( n_{0\downarrow} + n_{1\downarrow}) &  0 & 0 & 0 &  0 & 0 \\ 0 &  U ( n_{0\downarrow} + n_{1\uparrow}) & 0 & -t &  -t & 0 \\  0 &  0 & U ( n_{0\uparrow} + n_{1\downarrow}) & -t &  -t & 0 \\  0 &  -t & -t & U ( n_{0\downarrow} + n_{0\uparrow}) &  0 & 0 \\  0 &  -t & -t & 0 &  U ( n_{1\downarrow} + n_{1\uparrow}) & 0 \\ 0 &  0 & 0 & 0 &  0 & U ( n_{0\uparrow} + n_{1\uparrow}) ),
\label{eq:H_MF_hub_many_el_basis}
\end{equation}
\end{widetext}
where we removed the last term of eq.~(\ref{eq:ham_hubb_mf}) as in the MF-U case since it is only a constant shift of the Hamiltonian, leaving the eigen-vectors unchanged.

Similar to the MF-U case, the MF-MB Hamiltonian depends on the mean value of density operators that we compute self-consistently from the ground state of the preceding iteration, starting from an initial guess for the mean densities. As the ground state is given as a linear combination of the basis states given by eq.~(\ref{eq:many_el_basis}), the mean densities of electrons at site $i$ with spin $\sigma$ are computed by summing the square of the linear coefficients when the basis state contains an electron at site $i$ with spin $\sigma$. A new Hamiltonian is also computed iteratively, and then diagonalized to obtain to new mean densities, until self-consistency is achieved.

Numerically, the many-electron basis is encoded exactly in the same way as in ED, \textit{i.e.} using the integers $I$, $I_\uparrow$, and $I_\downarrow$. As in the ED case, the effect of creation, destruction, and density operators are implemented using simple binary operations on the binary representation of the three integers.

\subsection{Correspondence with MF-U}

We are now in a position to examine the links between MF-U and MF-MB. In this section, we explain and show with examples how the two methods are consistent. The MF-U method yields individual states (eigen-vectors of the Hamiltonian) for effectively independent particles, at given energies (eigen-energies of the Hamiltonian). The ground state of the system containing $N_{\rm el}$ is then formed by populating the $N_{\rm el}$ eigen-states with the lowest energies according to Pauli's principle. The total energy of the ground state is the sum of the individual energies. 

The counterpart (\textit{i.e.}, the ground state) in the MF-MB method is the lowest energy state which is a linear combination of the many-body basis states. The total energy of the ground state (containing $N_{\rm el}$ electrons) is the eigen-energy of the state, \textit{i.e.} the lowest eigen-energy of the MF-MB Hamiltonian.

We illustrate this with the two-site system at half-filling. We first introduce the change of basis from a localised basis to a bonding/anti-bonding basis in order to better understand the relation between MF-U and MF-MB results. In the single-electron picture, we define bonding (b) and anti-bonding (a) states for a given spin $\sigma$ as:
\begin{equation}
\begin{split}
    & \ket{b\sigma} = \frac{1}{\sqrt{2}} (\ket{0\sigma} + \ket{1\sigma}) = \frac{1}{\sqrt{2}} ( \hat{c}^\dagger_{0\sigma}  + \hat{c}^\dagger_{1\sigma}) \ket{\emptyset} {\rm~ and}\\
    & \ket{a\sigma} = \frac{1}{\sqrt{2}} (\ket{0\sigma} - \ket{1\sigma}) = \frac{1}{\sqrt{2}} ( \hat{c}^\dagger_{0\sigma}  - \hat{c}^\dagger_{1\sigma}) \ket{\emptyset},
\end{split}
\end{equation}
such that it is intuitive to define bonding and anti-bonding creation operators as:
\begin{equation}
    \begin{split}
        & \hat{c}^\dagger_{b\sigma} = \frac{1}{\sqrt{2}} ( \hat{c}^\dagger_{0\sigma}  + \hat{c}^\dagger_{1\sigma}){\rm~ and}  \\
        & \hat{c}^\dagger_{a\sigma} = \frac{1}{\sqrt{2}} ( \hat{c}^\dagger_{0\sigma}  - \hat{c}^\dagger_{1\sigma}).
    \end{split}
\end{equation}

We also define the many-body states in the bonding/anti-bonding basis by the successive applications of bonding and anti-bonding operators. As for the localised basis, it is important to choose a convention (\textit{fermionic sign}). We chose to define a positive sign state in our convention. We consider the states presenting the ordering of all spin-up (resp. -down) creation operators on the left (resp. right). Within the same spin part, a positive sign state orders the bonding and anti-bonding operators from left to right. For example, the many-body state containing two bonding states (of opposite spins) is expressed as
\begin{equation}
\begin{split}
    \ket{b\uparrow, b\downarrow}  & = \hat{c}^\dagger_{b\uparrow} \hat{c}^\dagger_{b\downarrow} \ket{\emptyset} \\
    & = \frac{1}{2} (\hat{c}^\dagger_{0\uparrow}  + \hat{c}^\dagger_{1\uparrow}) (\hat{c}^\dagger_{0\downarrow}  + \hat{c}^\dagger_{1\downarrow}) \ket{\emptyset} \\
    & = \frac{1}{2} ( \ket{\Phi_4} + \ket{\Phi_2} + \ket{\Phi_3} + \ket{\Phi_5}) .
\end{split}
\end{equation}

The eigen-states and eigen-energies for the MF approximation are listed in table~\ref{tab:two_site_MF_U} for the MF-U method and in table~\ref{tab:two_site_MF_MB} for the MF-MB method. The MF-U method yields four eigen-states at energies $E=-0.75t$ and $E=1.25t$. Each energy corresponds to doubly-degenerated states, due to spin. The single-particle eigen-states are the bonding and anti-bonding states. The MF-MB method gives six eigen-states with energies $E=-1.5t$ (non-degenerated), $E=0.5t$ (four times degenerated), and $E=2.5t$ (non degenerated). The second part of table~\ref{tab:two_site_MF_MB} indicates that the states formed by successive applications of bonding and anti-bonding operators are the eigen-states of the Hamiltonian. The ground state of the MF-MB ($E=-1.5t$) is $\ket{b\uparrow, b\downarrow}$. This state corresponds to the state formed by occupying the two lowest energy single-particle states of the MF-U method. Those states are $\ket{b\uparrow}$ and $\ket{b\downarrow}$, both with energies $E=-0.75t$. The energy of the ground state in the MF-MB method is then the sum of the individual energies of the two lowest energy states in MF-U. This illustrates that, for a given number of particles, the MF-U and MF-MB methods predict the same ground state with the same energy, as expected. The excited many-electron states computed in the MF-MB method (with energies $E=0.5t$ and $E=2.5t$) are the two-electron states formed by populating every other combinations of two single-particle states of MF-U (not the two lowest energy states).

The MF-MB thus predicts directly all the many-electron states that can be obtained from populating single-particle eigen-states from the MF-U method. In the $N_{\rm el}$ sector, a correspondence is therefore established and the two methods lead to consistent results.

\begin{table}
    \centering
    \begin{tabular}{c|c|c|c|c}
        Localised basis : & $\ket{0\uparrow}$ & $\ket{1\uparrow}$ & $\ket{0\downarrow}$ & $\ket{1\downarrow}$  \\
        \hline
        $E = -0.75t$ & $1/\sqrt{2}$  & $1/\sqrt{2}$ & 0 & 0 \\
        $E = -0.75t$ & 0 & 0 & $1/\sqrt{2}$  & $1/\sqrt{2} $\\
        $E = 1.25t$ & $1/\sqrt{2}$  & $-1/\sqrt{2}$ & 0 & 0 \\
        $E = 1.25t$ & 0 & 0 & $1/\sqrt{2}$  & $-1/\sqrt{2}$ \\
        \hline
        \\
        \hline
        Bonding/anti-bonding basis : & $\ket{b\uparrow}$ & $\ket{b\downarrow}$ & $\ket{a\uparrow}$ & $\ket{a\downarrow}$  \\
        \hline
        $E = -0.75t$ & 1 & 0 & 0 & 0 \\
        $E = -0.75t$ & 0 & 1 & 0 & 0 \\
        $E = 1.25t$ & 0 & 0 & 1 & 0 \\
        $E = 1.25t$ & 0 & 0 & 0 & 1
    \end{tabular}
    \caption{List of MF-U eigen-states of the two-site Hubbard system at half-filling with $U=0.5t$. There are four eigen-states, two pairs of degenerate states (spin degeneracy): one at $E=-0.75t$ and the other at $E=1.25t$. The first part of the table shows the eigen-states' coefficients in the single-electron basis constructed with localised basis states (see eq.~(\ref{eq:single_electron_basis})). The second part of the table gives the eigen-states' coefficients in the diagonal basis consisting of bonding and anti-bonding states.}
    \label{tab:two_site_MF_U}
\end{table}

\begin{table*}
    \centering
    \begin{tabular}{c|c|c|c|c|c|c}
        Localised basis :  & $\ket{\Phi_1} = \ket{\uparrow, \uparrow}$ & $\ket{\Phi_2} =\ket{\uparrow, \downarrow}$ & $\ket{\Phi_3} =\ket{\downarrow, \uparrow}$ & $\ket{\Phi_4} =\ket{\uparrow \downarrow, .}$ & $\ket{\Phi_5} =\ket{., \uparrow \downarrow}$    & $\ket{\Phi_6} =\ket{\downarrow, \downarrow}$ \\
        \hline
        $E = -1.5t$ & 0 & 1/2 & 1/2 & 1/2 & 1/2 & 0 \\
        $E = 0.5t$ & 0 & $-1/\sqrt{2}$ & $1/\sqrt{2}$ & 0 & 0 & 0 \\
        $E = 0.5t$ & 0 & 0 & 0 & $-1/\sqrt{2}$ & $1/\sqrt{2}$ & 0  \\
        $E = 0.5t$ & 1 & 0& 0& 0& 0& 0 \\
        $E = 0.5t$ &  0 & 0& 0& 0& 0& 1  \\
        $E = 2.5t$ & 0 & 1/2 & 1/2 & -1/2 & -1/2 & 0 \\
        \hline
        \\
        \hline
        Bonding/anti-bonding basis :  &  $\ket{b\uparrow, a\uparrow}$ & $\ket{b\uparrow, b\downarrow}$ & $\ket{b\uparrow, a\downarrow}$ & $\ket{a\uparrow, b\downarrow}$ & $\ket{a\uparrow,a \downarrow}$    & $\ket{b\downarrow, a\downarrow}$ \\
        \hline
        $E = -1.5t$ & 0 & 1 & 0 & 0 & 0 & 0 \\
        $E = 0.5t$ & 0 & 0 & 1 & 0 & 0 & 0 \\
        $E = 0.5t$ & 0 & 0 & 0 & 1 & 0 & 0  \\
        $E = 0.5t$ & 1 & 0& 0& 0& 0& 0 \\
        $E = 0.5t$ &  0 & 0& 0& 0& 0& 1  \\
        $E = 2.5t$ & 0 & 0 & 0 & 0 & 1 & 0
    \end{tabular}
    \caption{Table of the MF-MB eigen-vectors of the two-site Hubbard system for $U=0.5t$. There are six eigen-states with energies $E=-1.5t$, $E=0.5t$ (four times degenerated), and $E=2.5t$. The first part of the table shows the linear coefficients of the states in the many-body basis presented in sec.~\ref{sec:ED}) (localised basis). The second part of the table gives the eigen-states' coefficients in the bonding/anti-bonding basis, the basis constructed from the tensor product of the diagonal basis for the single-electron picture (MF-U).}
    \label{tab:two_site_MF_MB}
\end{table*}

\section{Density of states}

We now turn our attention to the notion of density of states in the MF-U and in the MF-MB approaches. Specifically, we highlight two possible interpretations of the DOS. This leads to different features in the DOS that will be discussed, such as the appearance of satellite peaks in the DOS that are usually attributed to \textit{beyond-MF} methods.

In the MF-U method, the DOS consists of Dirac delta-peaks at the energies of the single-particle states. The Fermi level is positioned between the highest-occupied and the lowest-unoccupied levels. From the MF-MB results, we can retrieve this DOS by adopting a picture based on differences between MF-MB eigen-energies. Indeed, starting from the ground state with $E=-1.5t$ and identifying the four states at $E=0.5t$ (see table~\ref{tab:two_site_MF_MB}) as states having exactly one electron in a bonding state and one electron in an anti-bonding state, we obtain the following system for single-particle state energies  

\[
\begin{cases}
E_{b\uparrow} + E_{b\downarrow} =-1.5t   & \text{(I)} \\
E_{b\uparrow} + E_{a\downarrow} =0.5t   & \text{(II)} \\
E_{a\uparrow} + E_{b\downarrow} =0.5t   & \text{(III)} \\
E_{b\uparrow} + E_{a \uparrow} =0.5t   & \text{(IV)} \\
E_{b\downarrow} + E_{a\downarrow} =0.5t   & \text{(V)} \\
E_{a\uparrow} + E_{a\downarrow} =2.5t   & \text{(VI)}. 
\end{cases}
\]

Using these six equations, we can find the single-particle state energies:
\begin{equation}
    \begin{split}
        E_{b\uparrow} = E_{b\downarrow} = -0.75t \\
        E_{a\uparrow} = E_{a\downarrow} = 1.25t.
    \end{split}
\end{equation}

This reasoning is somewhat inconvenient because it cannot be easily generalized to large systems. However, it presents the advantage of explaining how the MF-U DOS (see table~\ref{tab:two_site_MF_U}) can be recovered from the MF-MB method. However, we emphasize that this interpretation of the single-particle DOS does not match with the definition of the DOS in a many-electron basis, constructed from the Green's function (see eqs.~\ref{eq:many_body_greens_function} and~\ref{eq:DOS_MB_basis}). In the MB basis, the DOS explores, from the ground state in the $N_{\rm el}$ sector, all possible states in the $N_{\rm el}\pm 1$ sectors, corresponding to all possible ways of adding or removing one electron.

To adapt this definition of the DOS to MF-MB method, we need (as in the case of ED) to compute MF-MB results in the $N_{\rm el}$, $N_{\rm el}+1$, and $N_{\rm el}-1$ sectors independently and then compute the Green's function and the DOS from eqs.~\ref{eq:many_body_greens_function} and~\ref{eq:DOS_MB_basis}. The fundamental difference between this approach of the DOS and the approach of MF-U (with the correspondence with MF-MB explained before) is that:
\begin{itemize}
    \item in MF-U, all single-particle states are computed as interacting with the mean-field calculated from the ground state of the $N_{\rm el}$ sector;
    \item in MF-MB, peaks in the DOS corresponding to the addition (resp. removal) of one electron take into account the fact that when one electron is added (resp. removed), the states of the $N_{\rm el}+1$ (resp. $N_{\rm el}-1$) sector interact with the mean-field constructed from the ground state of the $N_{\rm el}+1$ (resp. $N_{\rm el}-1$) sector.
\end{itemize}

In MF-U, adding one electron to the ground state (containing $N_{\rm el}$) in the first unoccupied states will not result in the ground state of the system containing $N_{\rm el}+1$ electrons. Indeed, we would have to perform another MF-U calculation using a self-consistent procedure with the mean-field created from $N_{\rm el}+1$ electrons. This would even change the $N_{\rm el}$ lowest energy states from the previous MF-U calculation with $N_{\rm el}$ electrons.

In the MF-U method, the occupied states in the DOS have a clear physical meaning: they are the individual states composing the ground state. However, the interpretation of the unoccupied states is more difficult since they can't be attributed to physical states that will be occupied if we add an electron. 

In contrast, the definition of the DOS in MF-MB, in analogy with the ED technique, provides a clear physical interpretation of all the peaks: They represent all possible ways of adding or removing one electron from the $N_{\rm el}$-electron ground state.

We conclude this discussion with an important remark about the computation of the DOS in the MF-MB method. For some systems, starting from different initial conditions, it is possible to converge towards several different ground states, degenerated in energy. It seems unphysical to take into account only one of the ground-state solutions in the calculation of the DOS (eqs.~(\ref{eq:many_body_greens_function}) and~(\ref{eq:DOS_MB_basis})). At the same time, this reasoning might lead to the exclusion of states in some sectors. This idea is better explained by examining the example of the two-site system at half-filling. One needs to perform a MF-MB calculation in the one-electron and three-electron sectors to construct the MF-MB DOS. For illustration purposes, we only focus on the one-electron sector and on the removal part of the DOS. The one-electron sector eigen-states of the MF-MB method are given directly in the bonding/anti-bonding basis in table~\ref{tab:two_site_MF_MB_one_electron}. The single-particle eigen-states for the one-electron sector are the same as for the two-electron sector, but we observe that depending on the choice of the ground-state polarization (spin up or spin down), all other states are affected and spin-flipped.

\begin{table}
    \centering
    \begin{tabular}{c|c|c|c|c}
        Bonding/anti-bonding basis \\ spin-up ground state : & $\ket{b\uparrow}$ & $\ket{b\downarrow}$ & $\ket{a\uparrow}$ & $\ket{a\downarrow}$  \\
        \hline
        $E = -t$ & 1  & 0 & 0 & 0 \\
        $E = -0.75t$ & 0 & 1 & 0  & 0 \\
        $E = t$ & 0  & 0 & 1 & 0 \\
        $E = 1.25t$ & 0 & 0 & 0  & 1 \\
        \hline
        \\
        \hline
        Bonding/anti-bonding basis \\ spin-down ground state : & $\ket{b\uparrow}$ & $\ket{b\downarrow}$ & $\ket{a\uparrow}$ & $\ket{a\downarrow}$  \\
        \hline
        $E = -t$ & 0 & 1 & 0  & 0 \\
        $E = -0.75t$ & 1  & 0 & 0 & 0 \\
        $E = t$  & 0 & 0 & 0  & 1 \\
        $E = 1.25t$ & 0  & 0 & 1 & 0
    \end{tabular}
    \caption{List of eigen-states of the two-site Hubbard system at quarter-filling from the MF-U or MF-MB methods for $U=0.5t$. There are four eigen-states at energies $E=-t, -0.75t, t, $ and $1.25t$. The first (resp. second) part of the table shows the eigen-states' coefficients in the single-electron bonding/anti-bonding basis where the ground state was chosen to be spin up (resp. down).}
    \label{tab:two_site_MF_MB_one_electron}
\end{table}

When computing the Green's function (eq.~(\ref{eq:many_body_greens_function})), we have to consider all accessible states (in the $N_{\rm el}=1 $ sector) from the $N_{\rm el} =2$ ground state $\ket{b\uparrow,b\downarrow}$ (see table~\ref{tab:two_site_MF_MB}). For example, removing a spin-down electron from the state $\ket{b\uparrow,b\downarrow}$ results in the state $\ket{b\uparrow}$ that is present at different energies in the one-electron sector, depending on the ground-state polarization ($E=-t$ and $E=-0.75t$), see table~\ref{tab:two_site_MF_MB_one_electron}. It appears that only the state at $E=-t$ has to be taken into account since the other one is an unoccupied state when the ground state is spin-down polarised. Since we precisely chose to remove a spin-down electron from the $N_{\rm el}=2$ ground state, the one-electron ground state to be taken into account should be only the spin-up polarized one. The same reasoning holds for the removal of the spin-up electron. For the two-site system, there is a symmetry in the one-electron and three-electron sectors and thus in the addition and removal parts of the DOS.

\begin{figure}
\centering
    \includegraphics[width=8.5cm]{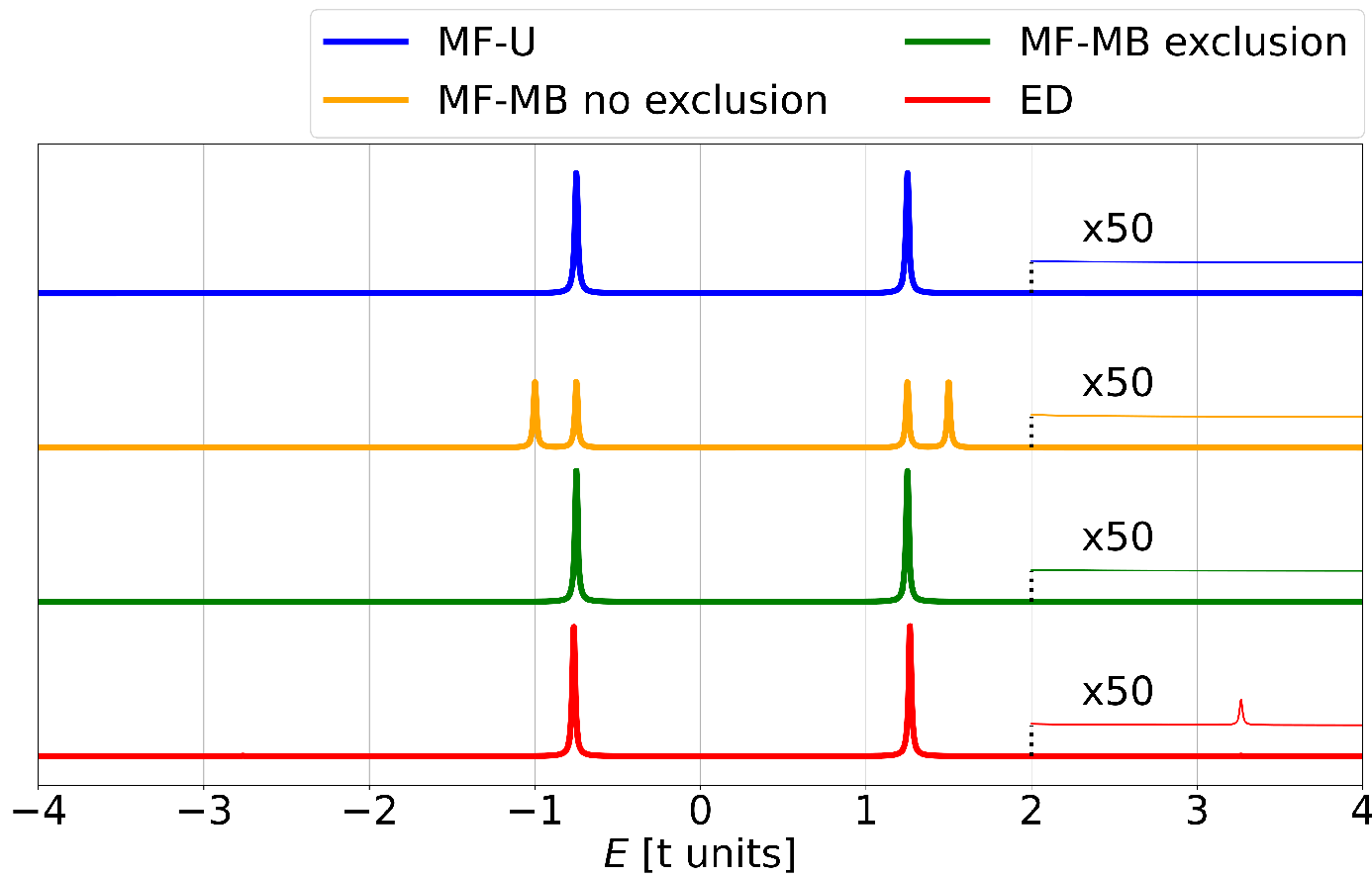}
\caption{Density of states of the two-site Hubbard system at half-filling with $U=0.5t$ for MF-U (blue), MF-MB without state exclusion in the $N_{\rm at}=1$ and $N_{\rm at}=3$ sectors (orange), MF-MB with state exclusion (green), and ED (red).}
\label{fig:DOS_Nat_2}
\end{figure}

The density of states computed in ED, MF-U, and MF-MB (with and without the exclusion of states in the $N_{\rm at}=1$ and $N_{\rm at}=3$ sectors) are given in fig.~\ref{fig:DOS_Nat_2}. This illustrates the process of state exclusion when comparing the two middle curves: the first one with no exclusion (second curve from the top at fig.~\ref{fig:DOS_Nat_2}) exhibits two peaks below the Fermi level, corresponding to a transition from the $N_{\rm el} =2$ ground state $\ket{b\uparrow,b\downarrow}$ to the states $\ket{b\uparrow}$ and $\ket{b\downarrow}$ for the two possible choices of ground state in the one-electron sector. In contrast, with the exclusion of states, the third curve from the top of fig.~\ref{fig:DOS_Nat_2} shows only one peak below the Fermi level, corresponding to the transition from the $N_{\rm el} =2$ ground state to the states $\ket{b\uparrow}$ and $\ket{b\downarrow}$ of $E=-t$ in the one-electron sector (the $E=-0.75t$ have been excluded). We note that for this specific case, we recover the MF-U result from the MF-MB with states exclusion but this is not the case in general. We also stress that ED predicts satellites (smaller peaks) for the two-site Hubbard model at half-filling, located at $E \simeq -2.77t$ and $E \simeq 3.27t$ (see Ref.~\onlinecite{strunck_combining_nodate, tomczak_proprietes_2007, honet_2021_2}). One satellite is visible in the insert of fig.~\ref{fig:DOS_Nat_2} but no satellite is present in either of the MF densities of states presented. We explore the presence of satellites in further detail in the next section.   

\section{Comparison between MF methods and ED}

We now illustrate the formalism described in this paper for the case of small atomic linear chains containing up to $3$ atomic sites, showcasing the main differences between the MF-U, MF-MB, and ED methods.

\subsection{MF-MB \textit{vs.} ED states and DOS: the two-site Hubbard system}
\label{sec:two-site}

We have shown above the results of the two-site problem for the MF-MB technique in the one-electron and two-electron sectors as well as the DOS in the half-filling case. Here, we start by recalling ED results that have been extensively studied before~\cite{romaniello_self-energy_2009, romaniello_beyond_2012, strunck_combining_nodate, tomczak_proprietes_2007,eder_introduction_2017}. We then describe the differences between MF-U, MF-MB, and ED states based on this simple model as well their consequences on the DOS. 

\begin{table}
    \centering
    \begin{tabular}{c|c|c|c|c}
        Bonding/anti-bonding basis : & $\ket{b\uparrow}$ & $\ket{b\downarrow}$ & $\ket{a\uparrow}$ & $\ket{a\downarrow}$  \\
        \hline
        $E = -t$ & 1 & 0 & 0  & 0 \\
        $E = -t$ & 0  & 1 & 0 & 0 \\
        $E = t$  & 0 & 0 & 1  & 0 \\
        $E = t$ & 0  & 0 & 0 & 1
    \end{tabular}
    \caption{List of ED eigen-states of the two-site Hubbard system at quarter-filling for $U=0.5t$. There are four eigen-states at energies $E=-t$ and $t$ (each doubly spin-degenerated).}
    \label{tab:two_site_ED_one_electron}
\end{table}

The eigen-vectors of the one-electron sector of the two-site Hubbard model with $U=0.5t$ are listed in table~\ref{tab:two_site_ED_one_electron} in the bonding/anti-bonding basis. They have to be compared to MF-MB eigen-states shown in table~\ref{tab:two_site_MF_MB_one_electron}. We observe that the main difference between the ED and MF-MB results is the lifting of spin-degeneracy in the MF-MB for both bonding and anti-bonding states. This can be understood as follows: in MF-MB, a specific polarized ground state is chosen, which leads to a mean-field interaction term that is only present for the opposite spin states. In ED, there is no interaction term since there is only one electron and no double occupation in the basis states.

\begin{table*}
    \centering
    \begin{tabular}{c|c|c|c|c|c|c}
        Localised basis :  &  $\ket{\uparrow, \uparrow}$ & $\ket{\uparrow, \downarrow}$ & $\ket{\downarrow, \uparrow}$ & $\ket{\uparrow \downarrow, .}$ & $\ket{., \uparrow \downarrow}$    & $\ket{\downarrow, \downarrow}$ \\
        \hline
        $E = -1.76556t$ & 0 & 0.5301 & 0.5301 & 0.467970 & 0.46797 & 0 \\
        $E = 0$ & 0 & $-1/\sqrt{2}$ & $1/\sqrt{2}$ & 0 & 0 & 0 \\
        $E = 0$ &  1 & 0& 0& 0& 0& 0 \\
        $E = 0$ & 0 & 0& 0& 0& 0& 1  \\
        $E = 0.5t$ & 0 & 0 & 0 & $-1/\sqrt{2}$ & $1/\sqrt{2}$ & 0  \\ 
        $E = 2.26556t$ & 0 & -0.46797 & -0.46797 & 0.5301 & 0.5301 & 0 \\
        \hline
        \\
        \hline
        Bonding/anti-bonding basis :  &  $\ket{b\uparrow, a\uparrow}$ & $\ket{b\uparrow, b\downarrow}$ & $\ket{b\uparrow, a\downarrow}$ & $\ket{a\uparrow, b\downarrow}$ & $\ket{a\uparrow,a \downarrow}$    & $\ket{b\downarrow, a\downarrow}$ \\
        \hline
        $E = -1.76556t$ & 0 & 0.99807 & 0 & 0 & -0.06214 & 0 \\
        $E = 0$ & 0 & 0 & $1/\sqrt{2}$ & $-1/\sqrt{2} $ & 0 & 0 \\
        $E = 0$ & 1 & 0 & 0& 0& 0& 0 \\
        $E = 0$ &  0 & 0& 0& 0& 0& 1  \\
        $E = 0.5t$ & 0 & 0 & $1/\sqrt{2}$ & $1/\sqrt{2}$ & 0 & 0  \\
        $E = 2.26556t$ & 0 & 0.06214 & 0 & 0 & 0.99807 & 0
    \end{tabular}
    \caption{List of the ED eigen-vectors of the two-site Hubbard system calculated for $U=0.5t$. There are six eigen-states with energies $E\simeq-1.76556t$, $E=0$ (three times degenerated), $E=0.5t$, and $E\simeq 2.26556t$. The first part of the table shows the linear coefficients of the states in the many-body basis of sec.~\ref{sec:ED}) (localised basis). The second part of the table gives the eigen-states' coefficients in the bonding/anti-bonding basis.}
    \label{tab:two_site_ED}
\end{table*}

The eigen-vectors of the two-site Hubbard model with $U=0.5t$ at half-filling are listed in table~\ref{tab:two_site_ED} in the localised and in the bonding/anti-bonding bases. These eigen-states have to be compared with MF-MB results of table~\ref{tab:two_site_MF_MB}. Note that if we want to compare the eigen-energies, we have to remember to take into account the constant term of eq.~(\ref{eq:ham_hubb_mf}) that consists in a constant shift in the energies. In the case of table~\ref{tab:two_site_MF_MB}, one has to apply a constant shift of $-0.25t$ for all eigen-energies. We observe three main effects of the MF approximation when comparing results from tables~\ref{tab:two_site_MF_MB} and~\ref{tab:two_site_ED}. The first MF effect is a well-known property: the MF approximation overestimates the ground-state energy. The difference between MF and ED ground-state energies is usually called the \textit{correlation energy}. The correlation energy $E_{\rm corr} = E_{\rm MF}-E_{\rm ED}$ depend on the parameter $U$ and, in our example, is given by $ E_{\rm corr}\simeq 0.01556t$. The second effect is that the three degenerate states at $E=0$ and the singly-degenerated state at $E=0.5t$ in ED are degenerate in MF at $E=0.5t$. This can be understood by noting that the basis states $\ket{\uparrow \downarrow,.}$ and $\ket{., \uparrow \downarrow}$ are treated in the same manner as basis states $\ket{\uparrow, \downarrow}$ and $\ket{\downarrow, \uparrow}$ in MF, whereas it is not the case in ED: the $E=0.5t$ singlet is formed by the basis states $\ket{\uparrow \downarrow,.}$ and $\ket{., \uparrow \downarrow}$ that are the only basis states introducing a Hubbard term since they induce a doubly-occupied site. This difference of treatment in MF and in ED is also responsible for the third observed difference: the MF ground state (at $E=-1.5t$) has equal weight for $\ket{\uparrow, \downarrow}$ and $\ket{\downarrow, \uparrow}$ than for $\ket{\uparrow \downarrow,.}$ and $\ket{., \uparrow \downarrow}$ whereas there is an asymmetry in the ED ground state due to the Hubbard term induced by the doubly-occupied site. The equal weight in MF results in the ground state being a product state (\textit{i.e.}, a Slater determinant) of single-electron eigen-states (bonding and anti-bonding states) and the ED ground states is a linear combination of Slater determinant of single-electron eigen-states. 

Because the ED ground state is a superposition of $\ket{b\uparrow, b\downarrow}$ and $\ket{a\uparrow, a\downarrow}$, the Green's function (eq.~(\ref{eq:many_body_greens_function})) features two poles for the part related to the $N_{\rm el }-1$-electron sector since the ground state couples with all the possible states of table~\ref{tab:two_site_ED_one_electron}. The same holds for the $N_{\rm el }+1$ electron sector part of the Green's function due to the symmetry of the two-site Hubbard model for $N_{\rm el}=1$ and $N_{\rm el}=3$. The coupling between the $N_{\rm el}=2$ ground state and the excited states of the $N_{\rm el}=1$ sector ($E=t$ states shown in table~\ref{tab:two_site_ED_one_electron}) results in the so-called \textit{satellite peaks} in the DOS of the half-filled two-site Hubbard system. A key feature of this peak is that it disappears at the $U=0$ limit: this can be explained since at $U=0$, the asymmetry between the two states $\ket{\uparrow, \downarrow}$ and $\ket{\downarrow, \uparrow}$ and the two states $\ket{\uparrow \downarrow,.}$ and $\ket{., \uparrow \downarrow}$ is canceled and the ground state is found to be a single Slater determinant of single-electron eigen-states $\ket{b\uparrow, b\downarrow}$. At the $U=0$ limit as well as in MF, the ground state is a pure product state of single-particle eigen-states such that it only couples to $\ket{b\uparrow}$ and $\ket{b\downarrow}$ listed in tables~\ref{tab:two_site_MF_MB_one_electron} and~\ref{tab:two_site_ED_one_electron}. Only one peak on each side of the Fermi level is observed as in the MF-MB exclusion curve of fig.~\ref{fig:DOS_Nat_2}.

We now turn to the quarter-filling DOS of the two-site Hubbard system. The ground state is either $\ket{b\uparrow}$ or $\ket{b\downarrow}$. Specifically, we chose the $N_{\rm at}=1$ ground state to be $\ket{b\uparrow}$. The Green's function of eq.~(\ref{eq:many_body_greens_function}) has poles corresponding to energy transitions between the single-particle ground state and all two-particle states listed in tables~\ref{tab:two_site_MF_MB} or~\ref{tab:two_site_ED} for MF and ED respectively as well as at energy transitions between the single-particle ground state and all zero-particle states. There exists only one zero-particle state: it is the vacuum state and has zero energy. 

\begin{figure}
\centering
    \includegraphics[width=8.5cm]{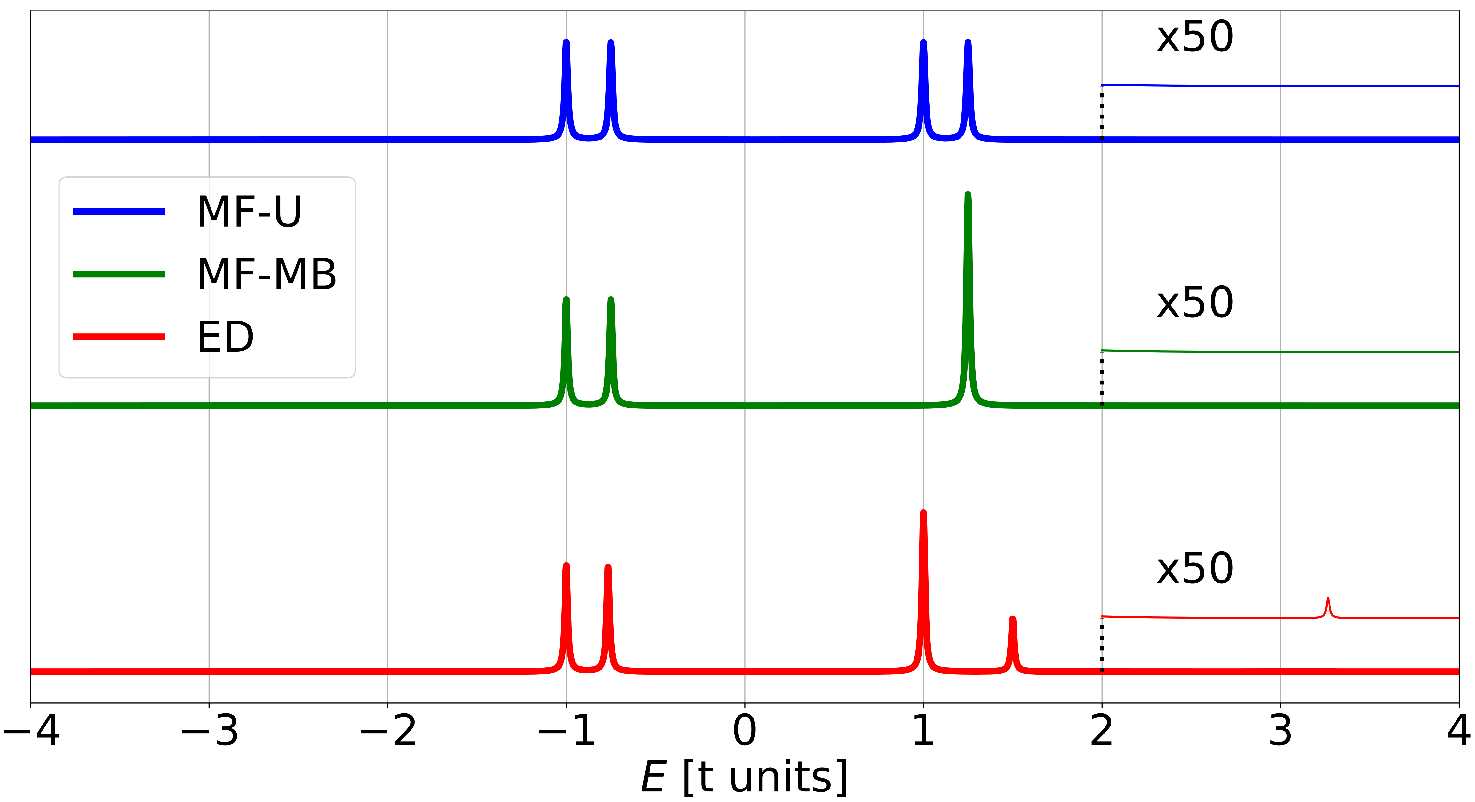}
\caption{Density of states of the two-site Hubbard system at quarter-filling with $U=0.5t$ for MF-U (blue), MF-MB (green), and ED (red).}
\label{fig:DOS_Nat_1}
\end{figure}

Figure~\ref{fig:DOS_Nat_1} shows the DOS of the quarter-filling two-site Hubbard system for $U=0.5t$ using the MF-U, MF-MB, and ED methods. Since there is only one electron in the system, the MF-U DOS exhibits peaks at the single-electron eigen-energies that are also found in MF-MB and given in table~\ref{tab:two_site_MF_MB_one_electron}. The lowest energy peak in MF-MB and ED methods are both removal peaks at the ground-state energy of the $N_{\rm el}=1$ sector of tables~\ref{tab:two_site_MF_MB_one_electron} and~\ref{tab:two_site_ED_one_electron}. 

The second lowest energy peak in MF-MB and ED corresponds to the coupling in the Green's function between the $N_{\rm el}=1$ and the $N_{\rm el}=2$ ground states. Since in the $N_{\rm el}=2$ sector the ED ground state is $\simeq 0.02t$ lower than the MF-MB one (correlation energy), the DOS peak is also found at a lower energy. In each case, it corresponds to the addition of a $\ket{b\downarrow}$ electron to the $N_{\rm el}=1$ ground state $\ket{b\uparrow}$. In MF-MB, the ground state is the only state in the $N_{\rm el}=2$ sector that contains that basis state $\ket{b\uparrow, b\downarrow}$. In ED, due to the asymmetry between states with and without doubly occupied states, the highest excited state also contains a part of the basis state $\ket{b\uparrow, b\downarrow}$. This is responsible for the presence of a satellite in the DOS, clearly visible in the insert of fig.~\ref{fig:DOS_Nat_1}. Alternatively, this satellite might also be interpreted as a contribution from the addition of a $\ket{b\downarrow}$ electron. The other consequence of this asymmetry is the fact that the second lowest-energy peak in ED does not integrate to 1, since it does not simply 
 represent a single-electron state.

Finally, the highest energy peak of the MF-MB DOS ($E=1.25 t$) originates from the coupling between the $N_{\rm el}=1$ ground state $\ket{b\uparrow}$ and two of the four states at $E=0.5t$ listed in table~\ref{tab:two_site_MF_MB}. The states $\ket{a\uparrow, b\downarrow}$ and $\ket{b\downarrow, a\downarrow}$ indeed cannot be reached from $\ket{b\uparrow}$ in the Green's function expression of eq.~\ref{eq:many_body_greens_function}. The observed peak in the DOS thus comes from the addition of $\ket{a\uparrow}$ and $\ket{a\downarrow}$ electrons. In MF-U, the addition of $\ket{a\uparrow}$ and $\ket{a\downarrow}$ electrons is non-degenerate because these unoccupied states are built by interacting with the mean-field produced by only one electron (the $\ket{b \uparrow}$ electron of the ground state) such that there is an interaction term for the addition of $\ket{a\downarrow}$ but not for the addition of $\ket{a\uparrow}$. In contrast, in MF-MB, the many-body states of the $N_{\rm el}=2$ sector are all constructed based on the mean-field produced by the $N_{\rm el}=2$ ground state ($\ket{b\uparrow, b \downarrow}$) that induces the same interaction term for the considered excited states. In ED, the peak is also split into two peaks in the DOS because of the lifting of degeneracy between $E=0$ and $E=0.5t$ energy states listed in table~\ref{tab:two_site_ED}. The $E=t$ peak in ED reflects the addition of a $\ket{a\uparrow}$ electron and the half of the addition of a $\ket{a\downarrow}$ electron. The other half is contained in the next peak in the DOS at $E=1.5t$. This splitting is due to the fact that adding a $\ket{a\downarrow}$ electron to a system containing a $\ket{b\uparrow}$ state involves contributions both from states with and without doubly-occupied sites translated into different interaction terms. The peak corresponding to the addition of a $\ket{a\uparrow}$ electron is not split since there is no interaction between two spin-up electrons.

\subsection{The emergence of satellites in MF-MB}

The MF approximation treats particles as independent particles interacting with a mean-field, neglecting all the correlation. It is often thought that satellites and quasi-particles are features that can only be observed when including correlation effects via approximations or exact treatment~\cite{reining_gw_2018,golze_gw_2019, romaniello_self-energy_2009, romaniello_beyond_2012,di_sabatino_reduced_2015}. Satellites have zero weight in the DOS for $U=0$ and an increasing weight when the interaction increases whereas quasi-particle peaks can be linked to particle peaks in the $U=0$ limit. Their weight might decrease as the interaction increases: a weight transfer from quasi-particle to satellite peaks then occurs~\cite{di_sabatino_reduced_2015}.

We now show that the MF-MB description presented here can also induce the emergence of satellites in linear chains containing $N_{\rm at} = 3$, showing that the satellite peaks are \textit{not} correlation peaks.

\begin{figure}
\centering
    \includegraphics[width=8.5cm]{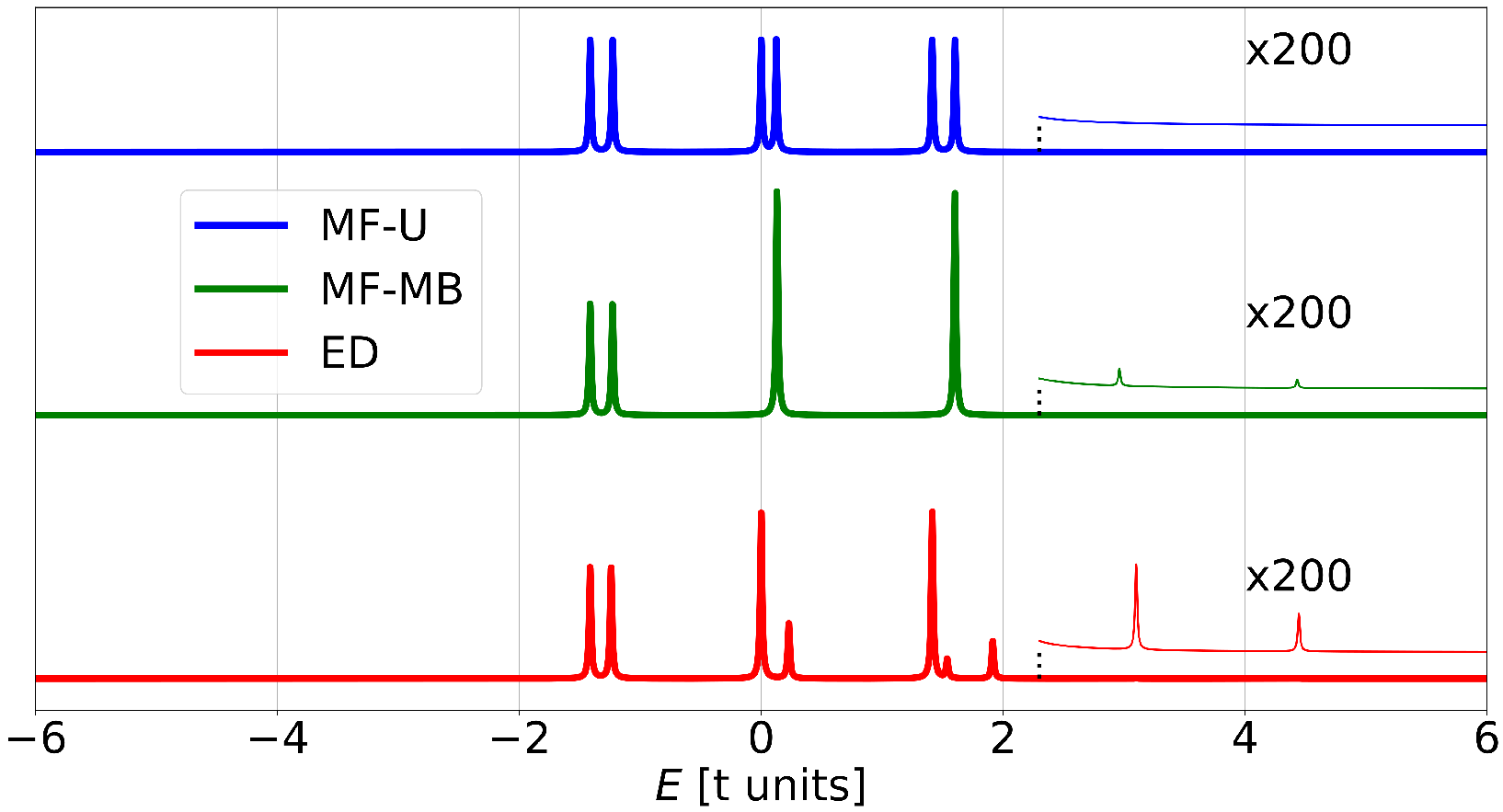}
\caption{Density of states of the three-site Hubbard system with one electron for $U=0.5t$ for MF-U (blue), MF-MB (green), and ED (red).}
\label{fig:DOS_Nat_3_Nel_1}
\end{figure}

The DOS for the $N_{\rm at}=3$ linear chain with one electron are shown in fig.~\ref{fig:DOS_Nat_3_Nel_1} for the MF-U, MF-MB, and ED methods. The DOS can be understood from the $N_{\rm el}=1$ and $N_{\rm el}=2$ eigen-states for MF-MB and ED given in the SI (tables 1 to 4). The general differences between the two methods have already been described for the two-site system in section~\ref{sec:two-site} and can be applied to the three-site system.

Similar to the two-site system at quarter-filling (see fig.~\ref{fig:DOS_Nat_1}), the two lowest peaks (\textit{i.e.}, the removal and the lowest-energy addition peaks) have approximately the same energies. The higher-energy pairs of peaks in MF-U look degenerated in MF-MB. However, the term \textit{degenerated} is not well-suited in the MF-MB case since, as for the ED DOS, each peak does not necessarily integrate to an integer value. This is related to the weight transfer from quasi-particle peaks to emergent satellites (at $E\simeq 2.96t$ and $E\simeq 4.43t$) that can be visualized on the zoomed-in MF-MB curve shown in fig.~\ref{fig:DOS_Nat_3_Nel_1}.

In ED, one observes that the two higher energy pairs of single-particle states in MF-U correspond to a principal peak that is more intense than the single-particle peaks, plus several smaller peaks located near the principal peaks and satellites that can be seen on the zoomed-in ED curve (see fig.~\ref{fig:DOS_Nat_3_Nel_1}). The small peaks near the principal ones are not reproduced in MF-MB for the same reason why the $E\simeq 1.5 t$ peak in the two-site system for ED is not present in the MF-MB (see fig.~\ref{fig:DOS_Nat_1}). Specifically, in MF-MB ($N_{\rm el}=2$ sector), there are fourfold-degenerated states that show a lifting of degeneracy and result in threefold-degenerated plus non-degenerated states (as discussed in section~\ref{sec:two-site} and can be seen in tables~\ref{tab:two_site_MF_MB} and~\ref{tab:two_site_ED}). The main observation for the three-site system with one electron is thus the emergence of satellites in MF-MB, that are also present in ED but absent in MF-U. The emergence of satellites in MF-MB is also observed for the three-site Hubbard system with filling up to half-filling (see SI). We expect satellites to be present also for larger systems since they contain an increasing number of accessible excited states.

\section{Conclusion}

In this work, we develop a method to compute the MF approximation in a many-body basis instead of a single-body basis. The MF approximation is a well-known and broadly used approximation in the single-body basis such that the interest of doing MF computations in a many-body basis, at much more expensive computational cost, might not be so straightforward. However, since ED techniques require a many-body basis treatment, the developed method allows for a better comparison between MF and ED.

We focus on the definition and the signification of the DOS and show how the usual MF approximation DOS can be found from MF-MB. We also highlight the possible ambiguity in a definition inspired by the ED DOS definition, coming from the use or not of Koopman's theorem (\textit{i.e.}, frozen single-particle states) in the prediction of the $N_{\rm el}\pm 1$ ground states. Attention is paid to the exclusion process, physically motivated by the lack of coherence in accounting for some of the different possible ground states, in the MF-MB DOS. The exclusion process is implemented manually in this work and an automated treatment of the process would be necessary for large-scale investigations.

We also observe that our MF-MB method induces satellites structure in the DOS. Satellites are usually seen as a sign of correlation since they appear in ED or approximations beyond MF (GW, T-matrix, DMFT,\ldots). We demonstrate in this work that satellites could appear in purely MF computed DOS, when adopting the ED-like definition. This result assumes the computation of states in $N_{\rm el} +1$ (resp. $N_{\rm el} -1$) sector using the mean-field based on the of the $N_{\rm el} +1$ (resp. $N_{\rm el} -1$) sector instead of using only one mean-field from the $N_{\rm el}$ sector.
 
The approach presented here contributes to a better understanding of the fundamental differences between MF and ED Hamiltonians and the associated energies and wave-functions. Our formulation includes Green's function expression that is the bridge between MF and many-body  corrections.  We therefore expect that this work could pave the way for developing other levels of approximation that introduce many-body correlations, such as GW or T-matrix, in the many-body basis in order to gain a deeper understanding of their mathematical and physical underpinnings. As further perspective, we suggest that revisiting and implementing known approximations from a refreshed point of view will deepen their understanding and lead to new approximations that would be deduced either numerically or from physical intuition. Specifically, we intend to explore the process of symmetry restoration of Refs.~\onlinecite{yannouleas_symmetry_2007} and~\onlinecite{sheikh_symmetry_2021} rewritten in MB basis. Finally, we posit that providing an in-depth and pedagogical analysis of a simple case of electronic correlation will help readers appreciate the difficulties associated with many-body electronic systems, especially since those are a cornerstone of quantum information and quantum computing developments.

\section*{Acknowledgements}

A.H. is a Research Fellow of the Fonds de la Recherche Scientifique - FNRS. This research used resources of the "Plateforme Technologique de Calcul Intensif (PTCI)" (\url{http://www.ptci.unamur.be}) located at the University of Namur, Belgium, and of the Université catholique de Louvain (CISM/UCL) which are supported by the F.R.S.-FNRS under the convention No. 2.5020.11. The PTCI and CISM are member of the "Consortium des Équipements de Calcul Intensif (CÉCI)" (\url{http://www.ceci-hpc.be}).


\bibliographystyle{ieeetr}
\bibliography{bibliography}

\end{document}


\preprint{APS/123-QED}

\title{Mean-field approximation of the Hubbard model expressed in a many-body basis: Supplementary information.}

\author{Antoine Honet}
\affiliation{%
Department of Physics and Namur Institute of Structured Materials, University of Namur, Rue de Bruxelles 51, 5000 Namur, Belgium
}%

\author{Luc Henrard}
\affiliation{%
Department of Physics and Namur Institute of Structured Materials, University of Namur, Rue de Bruxelles 51, 5000 Namur, Belgium
}%

\author{Vincent Meunier}%
\affiliation{%
Department of Engineering Science and Mechanics, The Pennsylvania State University, University Park, PA, USA
}%

\date{\today}

\maketitle


\author{Antoine Honet, Luc Henrard and Vincent Meunier}

\section{MF-MB and ED eigen-states for the three-site Hubbard system}

We consider here the three-site Hubbard system such that the single-particle localized basis states are given by $\ket{\downarrow, ., .}$, $\ket{., \downarrow, .}$, $\ket{., ., \downarrow}$, $\ket{\uparrow, ., .}$, $\ket{., \uparrow, .}$ and $\ket{., ., \uparrow}$. We also define the eigen-states (named $d$, $e$ and $f$) for the one-electron sector at $U=0$ case (i.e. \textit{tight-binding}):
\begin{equation}
    \begin{split}
        \ket{d\sigma} = \frac{1}{2}  \ket{\sigma, ., .} + \frac{1}{\sqrt{2}} \ket{., \sigma, .} +\frac{1}{2}  \ket{., ., \sigma} \\
        \ket{e\sigma} = \frac{-1}{\sqrt{2}}  \ket{\sigma, ., .} + \frac{1}{\sqrt{2}}  \ket{., ., \sigma} \\
        \ket{f\sigma} = \frac{-1}{2}  \ket{\sigma, ., .} + \frac{1}{\sqrt{2}} \ket{., \sigma, .} +\frac{-1}{2}  \ket{., ., \sigma}
    \end{split}
\end{equation}
where $\sigma$ stand for the spin and is eiter $\uparrow$ or $\downarrow$.

The MF-MB and ED eigen-states for the three-site Hubbard with one electron are given at tables~\ref{tab:three_site_MF_MB_one_electron} and~\ref{tab:three_site_ED_one_electron}, respectively. The MF-MB eigen-states are equivalant to MF-U eigen-states since there is only one electron in the system. The main difference between MF-MB and ED results is the lifting of spin degeneracy and the influence of the spin-polarized ground state on state with opposite spin in MF-MB. All eigen-states remain unchanged in ED when turning only the interaction (see table~\ref{tab:three_site_ED_one_electron} for $U=0.5t$), since there is no interaction for this case (no double occupied state). On the contrary, the eigen-states of spin opposite to the ground state spin in MF-MB are affected when increasing $U$ due to there interaction with the mean field based on the ground state. The eigen-state $\ket{e\sigma}$ remains however eigen-state for all $U$ values because the mean field is constant on all the localized sites where the state $\ket{e\sigma}$ is defined.

\begin{table*}[]
    \centering
    \begin{tabular}{c|c|c|c|c|c|c}
        Localised basis : & $\ket{\downarrow, ., .}$ & $\ket{., \downarrow, .}$ & $\ket{., ., \downarrow}$ & $\ket{\uparrow, ., .}$ & $\ket{., \uparrow, .}$ & $\ket{., ., \uparrow}$  \\
        \hline
        E = -1.41421 t & $1/2$  & $1/\sqrt{2}$ & 1/2 & 0 & 0 & 0 \\
        E = -1.22809t & 0 & 0 & 0  & 0.51092 & 0.69132 & 0.51092 \\
        E = 0 & $-1/\sqrt{2}$  & 0 & $1/\sqrt{2}$ & 0 & 0 & 0 \\
        E = 0.125t & 0 & 0 & 0 & $-1/\sqrt{2}$  & 0 & $1/\sqrt{2}$  \\
        E = 1.41421 t  & $-1/2$  & $1/\sqrt{2}$ & -1/2 & 0 & 0 & 0 \\
        E = 1.60309 t & 0 & 0 & 0 & -0.48884 & 0.72255 & -0.48884 \\
        \hline
        \\
        \hline
        $U=0$ diagonal basis : & $\ket{d\downarrow}$ & $\ket{e\downarrow}$ & $\ket{f\downarrow}$ & $\ket{d\uparrow}$ & $\ket{e\uparrow}$ & $\ket{f\uparrow}$  \\
        \hline
        E = -1.41421 t & 1  & 0 & 0 & 0 & 0 & 0 \\
        E = -1.22809t & 0 & 0 & 0  & 0.99976 & 0 & 0.02208 \\
        E = 0 & 0  & 1 & 0 & 0 & 0 & 0 \\
        E = 0.125t & 0 & 0 & 0 & 0  & 1 & 0  \\
        E = 1.41421 t  & 0  & 0 & 1 & 0 & 0 & 0 \\
        E = 1.60309 t & 0 & 0 & 0 & 0.02208 & 0 & -0.99976 \\
    \end{tabular}
    \caption{Table representing the eigen-states of the three-site Hubbard system with one electron in either the MF-U or the MF-MB methods and with $U=0.5t$. }
    \label{tab:three_site_MF_MB_one_electron}
\end{table*}

\begin{table*}[]
    \centering
    \begin{tabular}{c|c|c|c|c|c|c}
        Localised basis : & $\ket{\downarrow, ., .}$ & $\ket{., \downarrow, .}$ & $\ket{., ., \downarrow}$ & $\ket{\uparrow, ., .}$ & $\ket{., \uparrow, .}$ & $\ket{., ., \uparrow}$  \\
        \hline
        E = -1.41421 t & $1/2$  & $1/\sqrt{2}$ & 1/2 & 0 & 0 & 0 \\
        E = -1.41421 t & 0 & 0 & 0  & $1/2$  & $1/\sqrt{2}$ & 1/2 \\
        E = 0 & $-1/\sqrt{2}$  & 0 & $1/\sqrt{2}$ & 0 & 0 & 0 \\
        E = 0 & 0 & 0 & 0 & $-1/\sqrt{2}$  & 0 & $1/\sqrt{2}$  \\
        E = 1.41421 t  & $-1/2$  & $1/\sqrt{2}$ & -1/2 & 0 & 0 & 0 \\
        E = 1.41421 t & 0 & 0 & 0 & $-1/2$  & $1/\sqrt{2}$ & -1/2 \\
        \hline
        \\
        \hline
        $U=0$ diagonal basis : & $\ket{d\downarrow}$ & $\ket{e\downarrow}$ & $\ket{f\downarrow}$ & $\ket{d\uparrow}$ & $\ket{e\uparrow}$ & $\ket{f\uparrow}$  \\
        \hline
        E = -1.41421 t & 1  & 0 & 0 & 0 & 0 & 0 \\
        E = -1.41421 t & 0 & 0 & 0  & 1 & 0 & 0 \\
        E = 0 & 0  & 1 & 0 & 0 & 0 & 0 \\
        E = 0 & 0 & 0 & 0 & 0  & 1 & 0  \\
        E = 1.41421 t  & 0  & 0 & 1 & 0 & 0 & 0 \\
        E = 1.41421 t & 0 & 0 & 0 & 0 & 0 & 1 \\
    \end{tabular}
    \caption{Table representing the eigen-states of the three-site Hubbard system with one electron in ED and $U=0.5t$. }
    \label{tab:three_site_ED_one_electron}
\end{table*}


The eigen-states in the $N_{\rm el}=2$ sector in MF-MB and ED are given at tables~\ref{tab:three_site_MF_MB_two_electron} and~\ref{tab:three_site_ED_two_electrons}, respectively. The main difference is the lifting of degeneracy of four times degenerated states in MF-MB to three times degenerated states and non degenerated state in ED. We also observe the mixing of a pure product state of $U=0$ eigen-states ($\ket{e\uparrow, e\downarrow}$ at $E\simeq 0.26t$ in MF-MB) with other states in ED.

\section{DOS of for two and three electrons on three sites}

\begin{figure}
\centering
    \includegraphics[width=8.5cm]{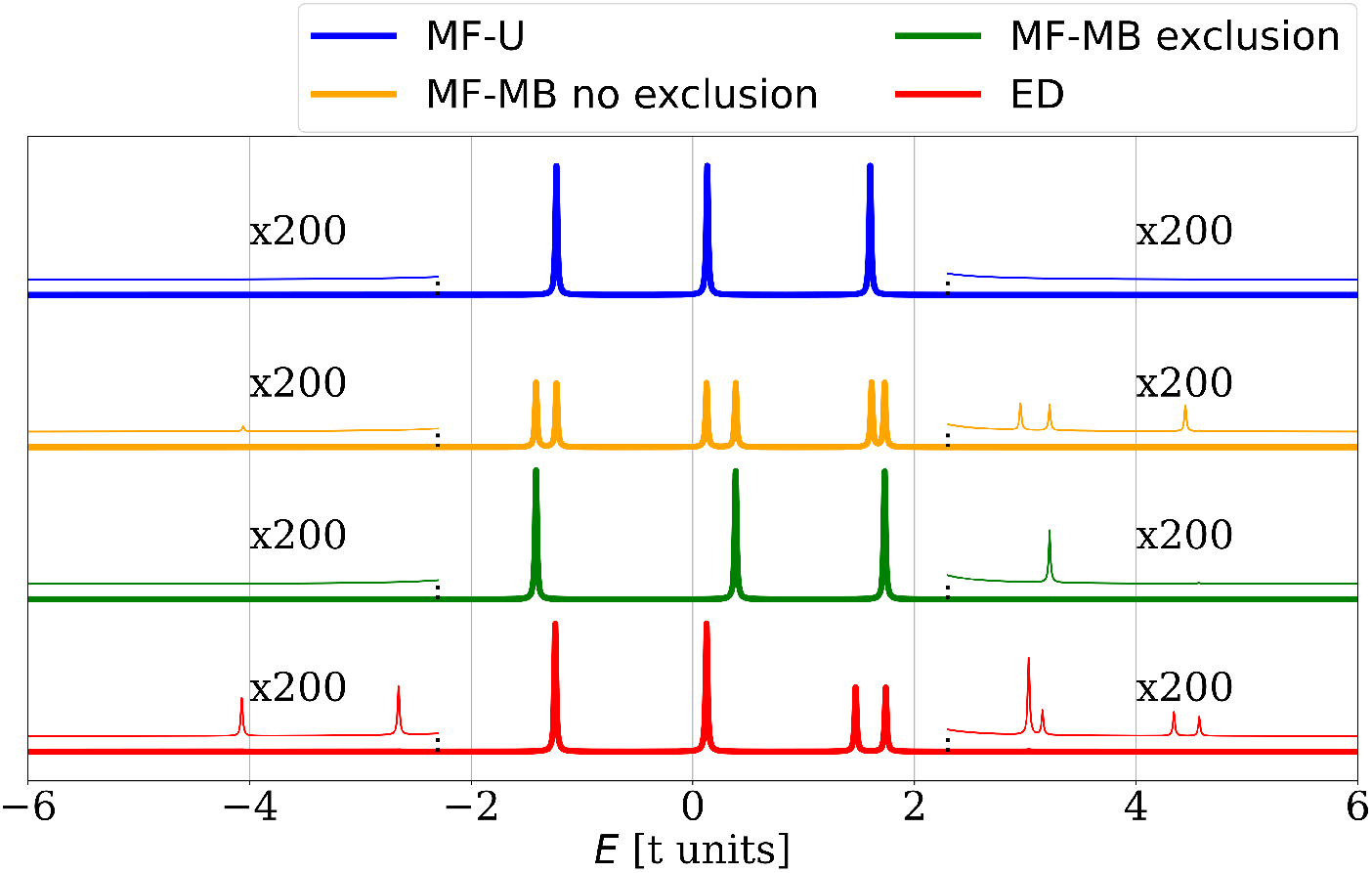}
\caption{Density of states of the three-site Hubbard system with two electrons with $U=0.5t$ for MF-U (blue), MF-MB (green) and ED (red).}
\label{fig:DOS_Nat_3_Nel_2}
\end{figure}

The DOS for the three-site Hubbard system filled with two electrons with $U=0.5t$ is given at fig.~\ref{fig:DOS_Nat_3_Nel_2}. Since the $N_{\rm el} = 2\pm 1$ sectors present polarized ground state, the exclusion mechanism explained in the main text is used such that MF-MB DOS is computed both with and without exclusion. We note the inclusion of satellites in the addition part of the spectrum. There are no satellites in the removal part since it correspond to exploring the $N_{\rm el}=1$ sector in which the eigen-states are single-particle eigen-states.

\begin{figure}
\centering
    \includegraphics[width=8.5cm]{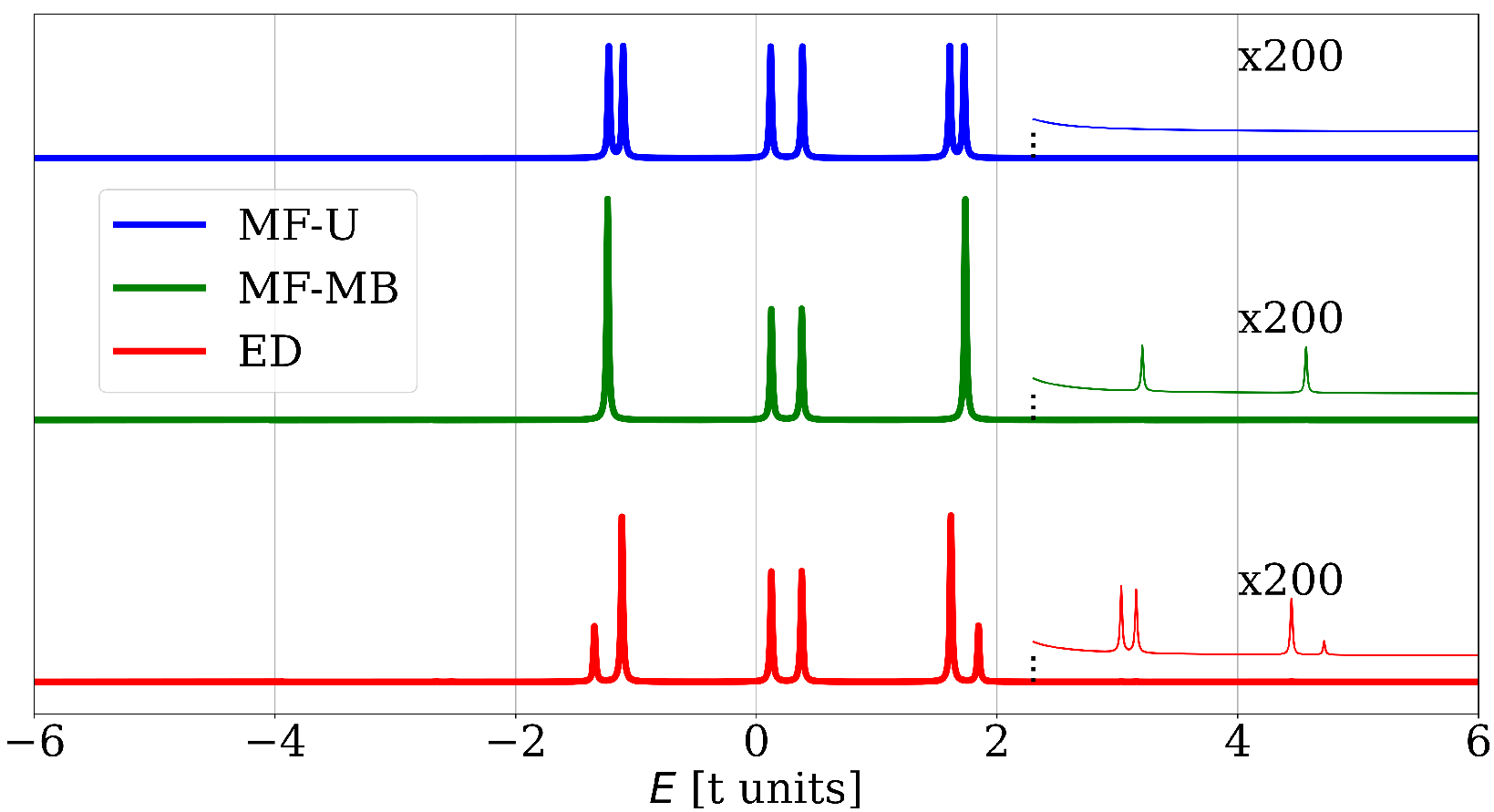}
\caption{Density of states of the three-site Hubbard system with three electrons with $U=0.5t$ for MF-U (blue), MF-MB (green) and ED (red).}
\label{fig:DOS_Nat_3_Nel_3}
\end{figure}

The DOS for the three-site Hubbard system filled with three electrons with $U=0.5t$ is given at fig.~\ref{fig:DOS_Nat_3_Nel_3}. As in the case of the three-site system with one electron exposed in the main text, we observe in MF-MB the pairing of states that are non-degenerated in MF-U. These states are split in ED and a part of the weight is transfered to satellites both in MF-MB and ED. In fig.~\ref{fig:DOS_Nat_3_Nel_3}, only the addition satellites are shown since there is a removal/addition symmetry.

\newpage

\begin{table*}[]
\addtolength{\tabcolsep}{-1pt}
    \centering
    \begin{tabular}{c|c|c|c|c|c|c|c|c|c|c|c|c|c|c|c}
        Loc. basis : & $\ket{., \downarrow, \downarrow}$ & $\ket{ \downarrow, ., \downarrow}$ & $\ket{ \downarrow, \downarrow,.}$ & $\ket{., ., \uparrow \downarrow}$ & $\ket{., \downarrow, \uparrow }$ & $\ket{ \downarrow, .,  \uparrow }$ & $\ket{., \uparrow, \downarrow }$ & $\ket{., \uparrow \downarrow, . }$ & $\ket{ \downarrow, \uparrow, . }$ &$\ket{\uparrow, . ,  \downarrow}$ & $\ket{\uparrow,   \downarrow, .}$ & $\ket{\uparrow   \downarrow, ., .}$ & $\ket{., \uparrow, \uparrow}$ & $\ket{ \uparrow, ., \uparrow}$ & $\ket{ \uparrow, \uparrow,.}$  \\
        \hline
        E = -2.46 t & & & & 0.25975 & 0.35328 & 0.25975 & 0.35328 & 0.4805 & 0.35328 & 0.25975 & 0.35328 & 0.25975 & & & \\
        E = -1.10 t & & & & 0.50935  & 0.3341 & -0.01804 & 0.35864 & & -0.35864 & 0.01804 & -0.33410 & -0.50934 & & &  \\
        E = -1.10 t & 0.49015  & 0.72076 & 0.49015 &  &  &  & & & & & & & & &  \\
        E = -1.10 t & & & & & & & & & & & & & 0.49015  & 0.72076 & 0.49015  \\
        E = -1.10 t  & & & & 0.01804 & 0.35864 & 0.50934 & -0.3341 & & 0.3341 & -0.50934 & -0.35864 & -0.01804 & & & \\
        E = 0.26 t & & & & $-1/2$  & & 1/2 & & & & 1/2 & & -1/2 & & & \\
        E = 0.37 t & & & & & & & & & & & & & $1/\sqrt{2}$ & & $-1/\sqrt{2}$ \\
        E = 0.37 t & $1/\sqrt{2}$ & & $-1/\sqrt{2}$ & & & &  & & & & & & & & \\
        E = 0.37 t & & & & 0.35324  & -0.02724 & 0.35324 & 0.01175 & -0.70648 &  -0.01175 & 0.35324 & -0.02724 & 0.35324 & & & \\
        E = 0.37 t & & & & -0.00547 & -0.49964 & -0.00547 & 0.50024  & 0.10943 & 0.50024 & -0.00547 & -0.49964 & -0.00547 & & & \\
        E = 1.73 t & -0.50966 & 0.69318 & -0.50966 & & & & & & & & & & & & \\
        E = 1.73 t & & & & & & & & & & & & & -0.50966 & 0.69318 & -0.50966 \\
        E = 1.73 t & & & & 0.48992  & -0.37131 & 0.0151 & -0.3491 & & 0.3491 & -0.0151 & 0.3713 & -0.48992 & & & \\
        E = 1.73 t & & & & -0.0151  & -0.3491 & 0.48992 & 0.37131 & & -0.37131 & -0.48992 &  0.3491 &  0.0151 & & & \\
        E = 3.20 t & & & & -0.24025  & 0.35328 & -0.24025 & 0.35328 & -0.5195 & 0.35328 & -0.24025 & 0.35328 & -0.24025 & & & \\
        \hline
        \\
        \hline
        $U=0$ \\ diag. basis : & $\ket{d\downarrow, e\downarrow}$ & $\ket{d\downarrow, f\downarrow}$ & $\ket{e\downarrow, f\downarrow}$ & $\ket{d\uparrow, d\downarrow}$ & $\ket{d\uparrow, f\downarrow}$ &
        $\ket{f\uparrow, d\downarrow}$ & $\ket{f\uparrow, f\downarrow}$  & $\ket{e\uparrow, e\downarrow}$ & $\ket{d\uparrow, e\downarrow}$ & $\ket{f\uparrow, e\downarrow}$ & $\ket{e\uparrow, d\downarrow}$ & $\ket{e\uparrow, f\downarrow}$ & $\ket{d\uparrow, e\uparrow}$ & $\ket{d\uparrow, f\uparrow}$ & $\ket{e\uparrow, f\uparrow}$ \\
        \hline
        E = -2.46 t & & & & 0.99962 & 0.0195 & 0.0195 & 0.0004 & & & & & & & & \\
        E = -1.10 t & & & & & & & & & -0.01125 & 0.70604 & 0.03881 & 0.70702 & & &  \\
        E = -1.10 t & -0.99981  & & 0.0195 &  &  &  & & & & & & & & &  \\
        E = -1.10 t & & & & & & & & & & & & & -0.99981  & & 0.0195  \\
        E = -1.10 t & & & & & & & & & 0.70702 & 0.03881 & -0.70604 & 0.01125 & & & \\
        E = 0.26 t & & & & & & & & 1 & & & & & & & \\
        E = 0.37 t & & 1 & &  & &  & & & & & & & & & \\
        E = 0.37 t & &  & &  & &  & & & & & & & & 1 & \\
        E = 0.37 t & & & & -0.02757 & 0.71744 & 0.69553 & 0.02757 & & & & & & & & \\
        E = 0.37 t & & & & 0.00043 & 0.69608 & -0.71796 & 0.00043 & & & & & & & & \\
        E = 1.73 t & 0.0195 & & 0.99981 & & & & & & & & & & & & \\
        E = 1.73 t & & & & & & & & & & & & & 0.0195 & & 0.99981 \\
        E = 1.73 t & & & & & & & & & 0.70621 & 0.008 & 0.70706 & -0.03557 & & &  \\
        E = 1.73 t & & & & & & & & & -0.03557 & 0.70706 & -0.008 & -0.70621 & & &  \\
        E = 3.20 t & & & & -0.00038 & 0.0195 & 0.0195 & -0.99962 & & & & & & & & \\
    \end{tabular}
    \caption{Table representing the eigen-states of the three-site Hubbard system with two electrons in MF-MB and $U=0.5t$. }
    \label{tab:three_site_MF_MB_two_electron}
\end{table*}


\begin{table*}[]
\addtolength{\tabcolsep}{-1pt}
    \centering
    \begin{tabular}{c|c|c|c|c|c|c|c|c|c|c|c|c|c|c|c}
        Loc. basis : & $\ket{., \downarrow, \downarrow}$ & $\ket{ \downarrow, ., \downarrow}$ & $\ket{ \downarrow, \downarrow,.}$ & $\ket{., ., \uparrow \downarrow}$ & $\ket{., \downarrow, \uparrow }$ & $\ket{ \downarrow, .,  \uparrow }$ & $\ket{., \uparrow, \downarrow }$ & $\ket{., \uparrow \downarrow, . }$ & $\ket{ \downarrow, \uparrow, . }$ &$\ket{\uparrow, . ,  \downarrow}$ & $\ket{\uparrow,   \downarrow, .}$ & $\ket{\uparrow   \downarrow, ., .}$ & $\ket{., \uparrow, \uparrow}$ & $\ket{ \uparrow, ., \uparrow}$ & $\ket{ \uparrow, \uparrow,.}$  \\
        \hline
        E = -2.66 t & & & & 0.23077 & 0.36404 & 0.27422 & 0.36404 & 0.46153 & 0.36404 & 0.27422 & 0.36404 & 0.23077 & & & \\
        E = -1.41 t & & & &  & 0.35355 & 0.5 & -0.35355 & & 0.35355 & -0.5 & -0.35355 & & & &  \\
        E = -1.41 t & 1/2  & $1/\sqrt{2}$ & 1/2 &  &  &  & & & & & & & & &  \\
        E = -1.41 t & & & & & & & & & & & & & 1/2  & $1/\sqrt{2}$ & 1/2 \\
        E = -1.19 t  & & & & 0.45440 & 0.38309 &  & 0.38309 & & -0.38309 & & -0.38309 & -0.45440 & & & \\
        E = 0 & $1/\sqrt{2}$ & & $-1/\sqrt{2}$ & & & &  & & & & & & & & \\
        E = 0 & & & & & & & & & & & & & $1/\sqrt{2}$ & & $-1/\sqrt{2}$ \\
        E = 0 & & & & & 1/2 & & -1/2 & & -1/2 & & 1/2 & & & & \\
        E = 0.12 t & & & & -0.20235  & -0.03801 & 0.61178 & -0.03801 & -0.40470 &  -0.03801 & 0.61178 & -0.03801 & -0.20235 & & & \\
        E = 0.5 t & & & & 0.57735 & & &  & -0.57735 & & & & 0.57735 & & & \\
        E = 1.41 t & -1/2 & $1/\sqrt{2}$ & -1/2 & & & & & & & & & & & & \\
        E = 1.41 t & & & & & & & & & & & & & -1/2 & $1/\sqrt{2}$ & -1/2 \\
        E = 1.41 t & & & &  & 0.35355 & -0.50 & -0.35355 & & 0.35355 & 0.50 &  -0.35355 &  & & &  \\
        E = 1.69 t & & & & 0.54177  & -0.32131 & & -0.32131 & & 0.32131 & & 0.32131 & -0.54177 & & & \\
        E = 3.03 t & & & & -0.26920  & 0.34064 & -0.22479 & -0.34064 & -0.53840 & 0.34064 & -0.22479  & 0.34064 & -0.26920 & & & \\
        \hline
        \\
        \hline
        $U=0$ \\ diag. basis : & $\ket{d\downarrow, e\downarrow}$ & $\ket{d\downarrow, f\downarrow}$ & $\ket{e\downarrow, f\downarrow}$ & $\ket{d\uparrow, d\downarrow}$ & $\ket{d\uparrow, f\downarrow}$ &
        $\ket{f\uparrow, d\downarrow}$ & $\ket{f\uparrow, f\downarrow}$  & $\ket{e\uparrow, e\downarrow}$ & $\ket{d\uparrow, e\downarrow}$ & $\ket{f\uparrow, e\downarrow}$ & $\ket{e\uparrow, d\downarrow}$ & $\ket{e\uparrow, f\downarrow}$ & $\ket{d\uparrow, e\uparrow}$ & $\ket{d\uparrow, f\uparrow}$ & $\ket{e\uparrow, f\uparrow}$ \\
        \hline
        E = -2.66 t & & & & 0.99808 & 0.02173 & 0.02173 & -0.03156 & 0.02173 & & & & & & & \\
        E = -1.41 t & & & & & & & & & $1/\sqrt{2}$ & & $-1/\sqrt{2}$ & & & &  \\
        E = -1.41 t & -1  & &  &  &  &  & & & & & & & & &  \\
        E = -1.41 t & & & & & & & & & & & & & -1 & &  \\
        E = -1.19 t & & & & & & & & & -0.70440 & -0.06178 & -0.70440 & -0.06178 & & & \\
        E = 0 & & 1 & &  & &  & & & & & & & & & \\
        E = 0 & &  & &  & &  & & & & & & & & 1 & \\
        E = 0 & & & & & $-1/\sqrt{2}$ & $1/\sqrt{2}$ & & & & & & & & & \\
        E = 0.12 t & & & & -0.05140 & 0.35330 & 0.35330 & 0.05612 & -0.81413 & & & & & & & \\
        E = 0.50 t & & & & & 0.57735 & 0.57735 & & 0.57735 & & & & & & & \\
        E = 1.41 t & & & 1 & & & & & & & & & & & & \\
        E = 1.41 t & & & & & & & & & & & & & & & 1 \\
        E = 1.41 t & & & & & & & & & & $-1/\sqrt{2}$ & & $1/\sqrt{2}$& & &  \\
        E = 1.69 t & & & & & & & & & -0.06178 & -0.70440 & -0.06178 & -0.70440 & & &  \\
        E = 3.03 t & & & & -0.03446 & 0.02221 & 0.02221 & -0.99792 & -0.04441 & & & & & & & \\
    \end{tabular}
    \caption{Table representing the eigen-states of the three-site Hubbard system with two electrons in ED and $U=0.5t$. }
    \label{tab:three_site_ED_two_electrons}
\end{table*}

\newpage
\turnpage

\begin{table*}[]
\addtolength{\tabcolsep}{-1pt}
    \centering
    \begin{tabular}{c|c|c|c|c|c|c|c|c|c|c|c|c|c|c|c|c|c|c|c|c|c|c}
        Loc. basis : & $\ket{\downarrow, \downarrow, \downarrow}$ & $\ket{ ., \downarrow, \uparrow \downarrow}$ & $\ket{ \downarrow, .,\uparrow \downarrow}$ & $\ket{., ., \uparrow \downarrow}$ & $\ket{\downarrow, \downarrow, \uparrow }$ & $\ket{  .,  \uparrow \downarrow, \downarrow }$ & $\ket{\downarrow, \uparrow, \downarrow }$ & $\ket{\downarrow, \uparrow \downarrow, . }$ & $\ket{ \uparrow,  \downarrow, \downarrow }$ &$\ket{\uparrow \downarrow, . ,  \downarrow}$ & $\ket{\uparrow,   \downarrow, .}$ & $\ket{\uparrow   \downarrow, \downarrow, .}$ & $\ket{., \uparrow, \uparrow \downarrow}$ & $\ket{ ., \uparrow\downarrow, \uparrow}$ & $\ket{\downarrow, \uparrow, \uparrow}$ & $\ket{ \uparrow, ., \uparrow \downarrow}$ & $\ket{ \uparrow, \downarrow, \uparrow }$ & $\ket{ \uparrow \downarrow,. , \uparrow }$  & $\ket{ \uparrow , \uparrow, \downarrow }$ & $\ket{ \uparrow , \uparrow \downarrow,. }$ & $\ket{ \uparrow \downarrow, \uparrow, . }$ & $\ket{ \uparrow , \uparrow, \uparrow }$ \\
        \hline
        E = -2.21 t & & 0.23728 & 0.3531 & 0.23728 & 0.3531 & 0.52544 & 0.3531 & 0.23728 & 0.3531 & 0.23728 & & & & & & & & & & \\
        E = -1.95 t & & & & & & & & & & & 0.26272 & 0.3531 & 0.26272 & 0.3531 & 0.47456 & 0.3531 & 0.26272 & 0.3531 &  0.26272 & \\
        E = -0.73 t & & -0.02544 & 0.34301 & 0.48645 & -0.38087 & & 0.38087 & -0.48645 & -0.34301 & 0.02544 & & & & & & & & & & \\
        E = -0.73 t & & -0.48645 & -0.38087 & -0.02544 & -0.34301 & & 0.34301 & 0.02544 & 0.38087 & 0.48645 & & & & & & & & & & \\
        E = -0.61 t  & & & & & & & & & & & 0.51242 & 0.33637 & -0.01188 & 0.35234 & & - 0.35234 & 0.01188 & -0.33637 &  -0.51242 & \\
        E = -0.61 t & & & & & & & & & & & 0.01188 & 0.35234 & 0.51242 & -0.33637 & & 0.33637 & -0.51242 & -0.35234 &  -0.01188 & \\
        E = 0.5 t & 1 & & & & & &  & & & & & & & & & & & & & \\
        E = 0.62 t & & -0.00666 & 0.49943 & -0.00666 & 0.50039 & 0.01332 & -0.50039 & -0.00666 & 0.49943 & -0.00666 & & & & & & & & & & \\
        E = 0.62 t & & -0.35303 & -0.03486 & -0.35303 & -0.016 & & 0.70607 & -0.016 & -0.03486 & -0.35303 & & & & & & & & & & \\
        E = 0.74 t & & & & & & & & & & & 0.5 & & -0.5 & & & & -0.5 & &  0.5 & \\
        E = 0.76 t &  & 0.5 & & -0.5 & & & & -0.5 & &  0.5 & & & & & & & & & & \\
        E = 0.88 t & & & & & & & & & & & 0.35303 & -0.03537 & 0.35303  & -0.0155 & -0.70605 &  -0.0155 & 0.35303  & -0.03537 &  0.35303 & \\
        E = 0.88 t & & & & & & & & & & & -0.00702 & -0.4994 & -0.00702  & 0.50041 & 0.01403 &  0.50041 & -0.00702  & -0.4994 &  -0.00702 & \\
        E = t & & & & & & & & & & & & & & & & & & & & & 1 \\
        E = 2.11 t & & 0.02053 & -0.35796 & 0.51215 & 0.33037 & & -0.33037 & -0.51215 & 0.35796 & -0.02053 & & & & & & & & & & \\
        E = 2.11 t & & 0.51215 & -0.33037 & -0.02053 & -0.35796 & & 0.35796 & 0.02053 & 0.33037 & -0.51215 & & & & & & & & & & \\
        E = 2.23 t & & & & & & & & & & & 0.4869 & -0.37302 & 0.01444 & -0.35153 & &  0.35153 & -0.01444  & 0.37302 &  -0.4869 & \\
        E = 2.23 t & & & & & & & & & & & -0.01444 & -0.35153 & 0.4869 & 0.37302 & &  -0.37302 & -0.4869  & 0.35153 &  0.01444 & \\
        E = 3.45 t & & -0.26272 & 0.35310 & -0.26272 & 0.35310 & -0.47456 & 0.35310 & -0.26272 & 0.35310 & -0.26272 & & & & & & & & & & \\
        E = 3.71 t & & & & & & & & & & & -0.23728 & 0.35310 & -0.23728 & 0.35310 & -0.52544 &  0.35310 & -0.23728  & 0.35310 &  -0.23728 & \\
        \hline
        \\
        \hline
        $U=0$ \\ diag. basis : & $\ket{d \downarrow, e \downarrow, f \downarrow}$ & $\ket{d \uparrow, d\downarrow,f\downarrow}$ & $\ket{d \uparrow, d\downarrow,e\downarrow}$ & $\ket{d \uparrow, e\downarrow,f\downarrow}$ & $\ket{f \uparrow, d\downarrow,f\downarrow}$ & $\ket{f \uparrow, d\downarrow,e\downarrow}$ & $\ket{f \uparrow, e\downarrow,f\downarrow}$ & $\ket{e \uparrow, d\downarrow,f\downarrow}$ & $\ket{e \uparrow, d\downarrow,e\downarrow}$ & $\ket{e \uparrow, e\downarrow,f\downarrow}$ & $\ket{d \uparrow, f\uparrow,d\downarrow}$ & $\ket{d \uparrow, f\uparrow,f\downarrow}$ & $\ket{d \uparrow, f\uparrow,e\downarrow}$ & $\ket{d \uparrow, e\uparrow,d\downarrow}$ & $\ket{d \uparrow, e\uparrow,f\downarrow}$  & $\ket{d \uparrow, e\uparrow,e\downarrow}$ & $\ket{e \uparrow, f\uparrow,d\downarrow}$ & $\ket{e \uparrow, f\uparrow,f\downarrow}$  & $\ket{e \uparrow, f\uparrow,e\downarrow}$ & $\ket{d \uparrow, f\uparrow,d\uparrow}$ \\
        \hline
        E = -2.21 t & & & 0.99935 & -0.02544 & & -0.02544 & 0.00065 & & & & & & & & & & & & & \\
        E = -1.95 t & & & & & & & & & & & & & & 0.99935 & 0.02544 & & 0.02544 & 0.00065 & & \\
        E = -0.73 t & & -0.74283 & & & 0.01891 & & & & 0.66899 & -0.01703 & & & & & & & & & & \\
        E = -0.73 t & & -0.66899 & & & 0.01703 & & & & -0.74283 & 0.01891 & & & & & & & & & & \\
        E = -0.61 t  & & & & & & & & & & & 0.6903 & 0.01757 & & & & 0.72308 & & &  0.01841 & \\
        E = -0.61 t & & & & & & & & & & & 0.72308 & 0.01841 & & & & -0.6903 & & &  -0.01757 & \\
        E = 0.5 t & 1 & & & & & &  & & & & & & & & & & & & & \\
        E = 0.62 t & & & 0.00068 & & 0.7203 & & -0.69366 & -0.00068 & & & & & & & & & & & & \\
        E = 0.62 t & & & 0.03597 & & 0.69273 & & 0.7194 & -0.03597 & & & & & & & & & & & & \\
        E = 0.74 t & & & & & & & & & & & & & -1 & & & & & & & \\
        E = 0.76 t & & & & & & & & & -1 & & & & & & & & & & & \\
        E = 0.88 t & & & & & & & & & & & & & & 0.03597 & -0.7201 & & -0.692 & -0.03597 & & \\
        E = 0.88 t & & & & & & & & & & & & & & -0.00071 & -0.69293 & & 0.721 & 0.00071 & & \\
        E = t & & & & & & & & & & & & & & & & & & & & & 1 \\
        E = 2.11 t & & -0.01726 & & & -0.678 & & & & 0.0187 & 0.73462 & & & & & & & & & & \\
        E = 2.11 t & & 0.0187 & & & 0.73462 & & & & 0.01726 & 0.678 & & & & & & & & & & \\
        E = 2.23 t & & & & & & & & & & & -0.01852 & 0.72753 & & & &  -0.01745 & & &  0.68561 & \\
        E = 2.23 t & & & & & & & & & & & -0.01745 & 0.68561 & & & & 0.01852 & & &  -0.72753 & \\
        E = 3.45 t & &  & 0.00065 & 0.02544 & & 0.02544 & 0.99935 & & & & & & & & & & & & & \\
        E = 3.71 t & & & & & & & & & & & & & & 0.00065 & -0.02544 & & -0.02544 & 0.99935 & & \\
    \end{tabular}
    \caption{Table representing the eigen-states of the three-site Hubbard system with three electrons in MF-MB and $U=0.5t$. }
    \label{tab:three_site_MF_MB_three_electrons}
\end{table*}

\newgeometry{left=0.1cm,bottom=0.1cm,top=0.1cm}
\restoregeometry

